# Damped Lyα Absorbers in Star-forming Galaxies at $z < 0.15$ Detected with the *Hubble Space Telescope* and Implications for Galaxy Evolution

Varsha P. Kulkarni,[1] David V. Bowen,[2] Lorrie A. Straka,[3] Donald G. York,[4, *]
Neeraj Gupta,[5] Pasquier Noterdaeme,[6, 7] and Raghunathan Srianand[5]

[1]*Department of Physics & Astronomy, University of South Carolina, Columbia, SC 29208, USA*
[2]*Department of Astrophysical Sciences, Princeton University, Princeton, NJ 08544, USA*
[3]*Sterrewacht Leiden, Leiden University, PO Box 9513, NL-2300 RA Leiden, the Netherlands*
[4]*Department of Astronomy & Astrophysics, University of Chicago, Chicago, IL 60637, USA*
[5]*Inter-University Centre for Astronomy and Astrophysics, Post Bag 4, Ganeshkhind 411007, Pune, India*
[6]*Institut d'Astrophysique de Paris, UMR7095, CNRS-UMPC, 98bis bd Arago, 75014 Paris, France*
[7]*Franco-Chilean Laboratory for Astronomy, IRL 3386, CNRS and Departamento de Astronomía, Universidad de Chile, Casilla 36-D, Santiago, Chile*

## ABSTRACT

We report *HST* COS spectroscopy of 10 quasars with foreground star-forming galaxies at $0.02 < z < 0.14$ within impact parameters of $\sim$1-7 kpc. We detect damped/sub-damped Lyα absorption in 100% of cases where no higher-redshift Lyman-limit systems extinguish the flux at the expected wavelength of Lyα absorption, obtaining the largest targeted sample of DLA/sub-DLAs in low-redshift galaxies. We present absorption measurements of neutral hydrogen and metals. Additionally, we present GBT 21-cm emission measurements for 5 of the galaxies (including 2 detections). Combining our sample with the literature, we construct a sample of 115 galaxies associated with DLA/sub-DLAs spanning $0 < z < 4.4$, and examine trends between gas and stellar properties, and with redshift. The H I column density is anti-correlated with impact parameter and stellar mass. More massive galaxies appear to have gas-rich regions out to larger distances. The specific SFR (sSFR) of absorbing galaxies increases with redshift and decreases with $M^*$, consistent with evolution of the star-formation main sequence (SFMS). However, $\sim$20% of absorbing galaxies lie below the SFMS, indicating that some DLA/sub-DLAs trace galaxies with longer-than-typical gas-depletion time-scales. Most DLA/sub-DLA galaxies with 21-cm emission have higher H I masses than typical galaxies with comparable $M^*$. High $M_{\rm HI}/M^*$ ratios and high sSFRs in DLA/sub-DLA galaxies with $M^* < 10^9 M_\odot$ suggest these galaxies may be gas-rich because of recent gas accretion rather than inefficient star formation. Our study demonstrates the power of absorption and emission studies of DLA/sub-DLA galaxies for extending galaxy-evolution studies to previously under-explored regimes of low $M^*$ and low SFR.

Corresponding author: Varsha P. Kulkarni
kulkarni@sc.edu

## 1. INTRODUCTION

An important challenge in current studies of galaxy evolution is understanding how galaxies interact with the intergalactic medium (IGM) via inflows and outflows of gas. Inflows of relatively cool metal-poor gas can provide fresh fuel for star formation. At the same time, outflows driven by feedback from supernovae and active galactic nuclei (AGN) can quench star formation. Both inflows onto galaxies and outflows from them pass through the circumgalactic medium (CGM). The CGM typically extends out to ∼300 kpc around the galaxies, and serves as the interface between the galaxies and the IGM. Indeed, the CGM could contain a substantial fraction of the so-far elusive (so-called "missing") baryons from the galaxy halos (e.g., Werk et al. 2014). Understanding how this CGM interacts with the galaxy's interstellar medium (ISM) is crucial in understanding the processes of star formation and feedback in galaxy evolution.

Unfortunately, it is difficult to observe the gas in and around distant galaxies directly in its own light owing to its low density. An excellent solution to this problem is to study this gas in *absorption* rather than emission, in the spectra of bright background sources such as quasars. Especially important for such studies are the strongest of the absorption systems, i.e., the damped Ly$\alpha$ (DLA) and the sub-damped Ly$\alpha$ (sub-DLA) systems. The DLAs have neutral hydrogen column densities $N_{\rm HI} \geq 2 \times 10^{20}$ cm$^{-2}$, while the sub-DLAs have $10^{19} \leq N_{\rm HI} < 2 \times 10^{20}$ cm$^{-2}$. Over ∼100,000 quasar absorption line systems, and $\gtrsim$12,000 candidate DLA/sub-DLAs have been discovered in the SDSS spectra of quasars (e.g., Noterdaeme et al. 2012b; York et al. 2022). DLAs and sub-DLAs provide the primary neutral gas reservoir for star formation (e.g., Péroux et al. 2005; Prochaska et al. 2005; Prochaska & Wolfe 2009; Noterdaeme et al. 2009; Noterdaeme et al. 2012b; Zafar et al. 2013; Popping et al. 2014). Furthermore, the DLAs and sub-DLAs offer the most reliable determinations of element abundances in distant galaxies (e.g., Pettini 1997; Kulkarni et al. 2005; Pettini 2011; Rafelski et al. 2012; Som et al. 2015; Quiret et al. 2016).

Given the high H I content of DLAs and sub-DLAs, their relevance for galaxy evolution studies is obvious. Indeed, DLAs and sub-DLAs are found to occur naturally in hydrodynamical simulations of structure formation, even out to a few hundred kpc from galaxy centers (e.g., Pontzen et al. 2008; Cen 2012; van de Voort12; Rahmati & Schaye 2014; Nelson et al. 2020). Unfortunately, the connections between DLA/sub-DLAs and galaxies still remain far from clear, because even after decades of attempts to image the galaxies producing the absorbers (e.g., Kulkarni et al. 2001; Rao et al. 2003; Straka et al. 2010; Peroux et al. 2011; Krogager et al. 2017), the galaxies associated with less than a few dozen DLA/sub-DLAs have been confirmed so far. As a result, a number of different scenarios exist for the origin of DLAs, ranging from galaxy disks to gas-rich dwarf galaxies to subgalactic clumps (e.g., Wolfe et al. 1986; York et al. 1986; Jimenez et al. 1999; Zwaan et al. 2008; Fynbo et al. 2013; Zafar et al. 2014; Cooke et al. 2015).

The primary observational difficulty in detecting the galaxies associated with DLA/sub-DLAs in the continuum is that the absorbing galaxies are faint compared to the background quasars, and

* Enrico Fermi Institute, University of Chicago, Chicago, IL 60637, USA

often at small angular separations from the quasars. This makes it difficult to reliably determine impact parameters, morphologies, sizes, structures, and star formation rates (SFRs) for the galaxies associated with the absorbers. These problems are especially challenging at high redshifts due to the $(1 + z)^{-4}$ dimming of surface brightness. A small sample of DLAs at $z>2$ has been found to have host galaxies with disk-like properties (e.g., Djorgovski et al. 1996; Møller et al. 2002). On the other hand, based on kinematic similarities with the Local Group dwarf population, it has been suggested that the general $z\sim3$ DLA population arises in metal-poor dwarf galaxies (Cooke et al. 2015). It has also been suggested that DLAs at $z\sim3$ are similar to the outskirts of compact Lyman Break Galaxies (LBGs) with much lower star formation efficiencies than observed in local galaxies (e.g., Rafelski et al. 2011). Indeed, deep emission-line imaging searches indicate that most high-$z$ DLAs have low SFRs (e.g., Kulkarni et al. 2000; Kulkarni et al. 2001; Kulkarni et al. 2006; Christensen et al. 2009; Peroux et al. 2011).

The problems in detecting galaxies associated with DLA/sub-DLAs are easier to address at $z\lesssim1$, where galaxy imaging is much easier. Indeed galaxies associated with some DLAs at $z<1$ have been detected in broad-band imaging studies (e.g. Le Brun et al. 1997; Bergeron et al. 1991; Rao et al. 2003) and in integral field spectroscopy (IFS; e.g., Peroux et al. 2011; Péroux et al. 2014; Péroux et al. 2019; Hamanowicz et al. 2020). Such studies suggest that even at $z<1$, a substantial fraction of the galaxies associated with DLAs are dwarf galaxies. Indeed, ∼50% of low-$z$ DLAs appear to arise in galaxies with $L<0.6L_*$; by contrast observations of a few sub-DLAs at $z<0.7$ show them all to be associated with $L>0.6L_*$ disk-dominated galaxies (e.g., Zwaan et al. 2005). We note, however, that such comparisons are complicated by the detections of multiple galaxies in the fields of some absorbers (e.g., Kacprzak et al. 2010; Péroux et al. 2017; Péroux et al. 2019; Hamanowicz et al. 2020).

Despite these successes, even at $z\sim1$, galaxies associated with some DLAs remain undetected. Furthermore, even for galaxies that are detected, it can be difficult to determine the detailed stellar properties for comparison with the gas properties. The IFS studies often offer no measurement of the continuum. Samples of galaxies at redshifts $\lesssim0.1$ can be probed at a much higher sensitivity that can uncover low surface brightness features, and are thus ideal for investigating the galaxy-CGM connection. An additional advantage is that at such low redshifts, the H I gas can also be detected directly through its 21 cm line emission. Since the atomic gas component of galaxies extends much farther out than the stellar component and is more affected by environmental processes, these observations can provide additional inputs to understand the origins of the gas producing DLA/subDLAs (e.g., Carilli & van Gorkom 1992; Gupta et al. 2010; Dutta et al. 2017). With this in mind, we have started targeting previously imaged galaxies at $z\lesssim0.1$ with background quasars at small angular separations, and examining absorption lines at the redshifts of these galaxies in the quasar spectra. Recently, studies using the Sloan Digital Sky Survey (SDSS; York et al. 2000) have revealed such a sample of low-$z$ galaxies on top of quasars (GOTOQs) (Borthakur et al. 2010; Noterdaeme et al. 2010; York et al. 2012; Straka et al. 2013; Straka et al. 2015; Joshi et al. 2017b).

GOTOQs are galaxies intervening with a background quasar for which part or all of the galaxy falls within the fibers of the SDSS spectrographs [with a diameter of 3″ for the SDSS spectrograph and 2″ for the Baryon Oscillation Spectroscopic Survey (BOSS) spectrograph]. Star-forming galaxies in the redshift range $0<z<0.40$ can leave their imprints in optical quasar spectra in the form of nine strong, narrow galactic nebular emission lines ([O II] $\lambda3727$, H$\beta$, [O III] $\lambda\lambda4960,5008$, [N II] $\lambda 6550$, H$\alpha$, [N II] $\lambda 6585$, [S II] $\lambda\lambda 6718,6733$). Thus, GOTOQs are star-forming galaxies that are detectable in spectral absorption as well as emission, and are therefore well-suited for studying the environment of quasar absorption systems. For galaxies with $0.40<z<0.80$, optical spectra can cover the emission lines of [O II], H$\beta$, [O III], detection of which, together with Mg II absorption lines, can be used to confirm the galaxies. An automated search for these lines in over 100,000 SDSS quasar spectra led to a catalog of $\sim$200 star-forming galaxies detected in these emission lines.

To investigate whether GOTOQs produce DLA/sub-DLAs, it is essential to obtain UV spectroscopy of the background quasars covering the Ly$\alpha$ absorption lines at the galaxy redshifts. Here we report results of *Hubble Space Telescope (HST)* Cosmic Origins Spectrograph (COS) observations of 10 star-forming galaxies at $z<0.14$. The SDSS spectra of the 10 quasars show absorption features of Ca II H, K and Na I D1, D2 at the redshifts of these galaxies. Moreover, superposed on these quasar spectra are narrow nebular emission lines from the foreground galaxies that fall within the same SDSS spectral fiber as the quasar. The COS FUV spectra allow us to measure the Ly$\alpha$ absorption associated with these galaxies, providing a direct constraint on the incidence of DLA/sub-DLAs in galaxies of these morphologies and environments. Furthermore, the COS spectra allow us to make approximate assessments of the metal line strengths, and to broadly search for any correlations between the gas properties inferred from the absorption features and the continuum and nebular emission properties of the galaxies. We also complement the SDSS and *HST* COS observations with H I 21-cm emission line observations of some of the galaxies to further understand the origin of absorbing gas with respect to these galaxies.

This paper is organized as follows: Section 2 details our sample selection and observations. Section 3 discusses the results of this study, including a summary of the characteristics of the individual objects in our sample in section 3.1. Section 4 describes the results of our search for the H I 21-cm emission line using the Green Bank Telescope (GBT) in several of the sample galaxies. Section 5 compares our results with those from the literature. Finally, Section 6 summarizes our conclusions. Throughout this paper, we adopt the concordance cosmology ($\Lambda$CDM; $\Omega_m$=0.3, $\Omega_\Lambda$=0.7, and H$_0$=70 km s$^{-1}$ Mpc$^{-1}$).

## 2. OBSERVATIONS AND DATA REDUCTION

### 2.1. *Sample Selection*

The SDSS images of each GOTOQ field in five optical broadband filters (*ugriz*) allow us to potentially resolve the morphology and geometry of these low-redshift galaxies. In 70% of cases in the parent sample (103 total), we detect a galaxy in the foreground of the quasar responsible for the nebular emission. The remaining 30% of cases do not have imaging detections of galaxies and are likely extremely faint dwarf galaxies within the quasar point spread function (PSF).

Our *HST* COS targets consisted of 10 GOTOQs selected from this larger sample. To construct our *HST* sample, we first selected quasars from the parent sample that have GALEX FUV flux $>30\mu$Jy in order to provide an adequate S/N in a reasonable number of *HST* orbits, and have redshifts $z_{quasar}<1.5$ to minimize the chances of intercepting an optically thick absorber along the line of sight which would extinguish the flux below its Lyman limit at the wavelengths where Ly$\alpha$ from the GOTOQ would be expected. These criteria ruled out all but 13 of the quasars in our parent sample. Our only other selection criterion was that the galaxy redshift be such that the Ly$\alpha$ line fell in a clean part of the COS grating, free of potential geo-coronal emission lines. These criteria resulted in a subset of ten GOTOQs suitable for COS observations of the H I Ly$\alpha$ absorption lines.

The left panels of Figure 1 and Figure 2 show the Dark Energy Camera Legacy Survey[1] (DECaLS; Dey et al. 2019) image thumbnails of our targets, centered on the background quasars. Figures 3 and 4 show the SDSS spectra of all of our target quasars, zoomed into the wavelength ranges containing the detected nebular emission lines produced by the foreground galaxies. Also shown are regions covering the Ca II H and K and Na I D absorption lines, although in most of these ten sight lines these absorption lines are elusive.

Our ten GOTOQs have redshifts $z<0.15$ and lie within impact parameters $\rho \lesssim 7$ kpc from the quasar lines of sight. They have active star formation with SFRs (estimated from $H\alpha$ emission) in the range of 0.01-0.26 $M_\odot$ yr$^{-1}$. (We note that the SFRs could be higher than these estimates; we discuss the SFRs more in sec. 2.2.2.) These galaxies range in luminosities from 0.02 $L^*$ to 2.86 $L^*$, thus covering both dwarf and non-dwarf galaxies, taking $L \lesssim 0.1 L^*$ as the definition of a dwarf galaxy (Cimatti et al. 2020). Some of these galaxies also exhibit Ca II H, K or Na I D1, D2 absorption lines in the SDSS spectra of the background quasars. The SDSS data provide the emission line metallicities, dust reddening, and five-band optical photometry for each of these galaxies. As mentioned above, the quasar sight lines are at close enough impact parameters from the galaxy centers to potentially probe the inner regions of the galaxies, as can be seen in the left panels of Figure 1 and Figure 2 . Table 1 lists our targets and their characteristics. The galaxy magnitudes are in the AB system (Oke et al. 1983) and have been corrected for foreground Milky Way extinction (Straka et al. 2015).

## 2.2. *HST/COS Data*

HST COS observations were obtained under program ID 14137 (PI: Straka). The FUV G140L grating was used with the central wavelength setting 1105 Å for seven quasar sight lines, and 1280 Å for the remaining three sight lines. Each quasar was observed for two orbits for total integration times ranging from 86 to 95 minutes. The foreground galaxies caused no problems with target acquisitions. All four focal plane positions (FP-POS) were used (2 positions per orbit) to help mitigate fixed-pattern noise. For two of the ten quasars observed, no flux was detected in the region of interest due to the presence of a previously unknown Lyman-limit system (LLS) at a higher redshift than that of the DLA of interest. For the remaining eight quasars, the COS spectra had adequate S/N. The G140L grating provides a spectral resolution of $\sim$1500 (i.e., a velocity resolution of $\sim$200 km s$^{-1}$) at $\sim$1350 Å. This resolution is adequate for detecting broad absorption lines, such as the Ly$\alpha$ lines in the DLAs/sub-DLAs hosted by the GOTOQs, and also enables approximate measurements of metal absorption lines (even if unresolved) arising in these galaxies.

### 2.2.1. *HST COS Spectral Analysis*

Each *HST* COS exposure was reduced and extracted with the Space Telescope Science Data Analysis System (STSDAS) COS pipeline. The individual extracted 1-dimensional spectra were co-added. The quasar continuum was fitted, and absorption line measurements were performed on the continuum-normalized spectra. In order to derive an H I column density and absorption velocity from the Ly$\alpha$ line, we first normalized the continuum of the quasar flux by fitting Legendre polynomials to wavelength ranges free from emission or absorption features (e.g. Sembach & Savage 1992). While establishing a best-fit continuum to the data, this method also generates $\pm 1\sigma$ different fits, which we refer to as "upper" and "lower" error envelopes which can also be used to normalize the data. We constructed theoretical Ly$\alpha$ Voigt profiles with initial estimates of the H I column densities $N_{\rm HI}$, velocity offsets $\Delta v$ relative to the galaxy redshifts (the latter defined by the nebular emission lines), and Doppler parameters, and varied these to fit the Ly$\alpha$ absorption in the spectrum normalized by the best fit continuum, as well as by the upper and lower envelopes (Bowen et al 2008, 2016). Theoretical COS G140L line spread functions were convolved with the Voigt line profiles, but in most cases the Ly$\alpha$ absorption observed was so broad and the H I column density so high that both the LSF widths and the Doppler parameters were too small to make an appreciable difference to the final estimates of the column densities.

---

[1] https://www.legacysurvey.org/decamls/

Values of $N_{\rm HI}$ and $\Delta v$ are given in Table 3. Errors in $N_{\rm HI}$ come from the difference between the values of $N_{\rm HI}$ derived from the upper and lower envelope fits to the continuum, and the best fit, and are significantly larger than any errors arising from Poisson statistics in the data alone. The right panels in Figure 1 and Figure 2 show the best-fitting Voigt profiles and $\pm 1\sigma$ profiles obtained for each of the absorbers. Seven of the eight absorbers are found to be DLAs, and one is found to be a sub-DLA.

Table 2 lists the metal line measurements detected at $\gtrsim 3\sigma$ level performed using the program SPECP[2] on the COS spectra for the eight absorbers for which the H I Ly$\alpha$ is covered. While the spectral resolution of our COS spectra is not adequate to obtain reliable metal column densities, the equivalent widths of the metal lines are well determined and are expected to be reliable. The quasar continuum was first fitted globally with a cubic spline polynomial (typically of order 1 to 2) using the IRAF task CONTINUUM. The continuum placement was refined locally near each line in SPECP before measuring the equivalent widths. For each measured feature, Table 2 also lists the 1 $\sigma$ uncertainty in the equivalent width due to photon noise, the 1 $\sigma$ uncertainty due to the continuum measurement uncertainty, the combined 1 $\sigma$ uncertainty (obtained by adding in quadrature the photon noise and continuum measurement uncertainties), and the significance of the feature.

### 2.2.2. *SDSS Measurements*

The SDSS fiber of 3″ diameter allows us to estimate emission line properties of the galaxy in approximately the same area as the COS circular aperture (2.5″ in diameter). Below we review the spectral and photometric measurements made for these galaxies from SDSS data, but refer the reader to our past papers (York et al. 2012; Straka et al. 2013; Straka et al. 2015) for more detailed discussion on the topic.

Detections of [O II] and H$\alpha$ emission lines allow us to estimate the SFR following the standard prescriptions (Kennicutt 1998):

$$SFR_{\rm [OII]}\ (M_\odot\ {\rm yr}^{-1}) = 1.4 \times 10^{-41}\ L_{\rm [OII]}\ ({\rm ergs\ s}^{-1}) \qquad (1)$$

$$SFR_{\rm H\alpha}\ (M_\odot\ {\rm yr}^{-1}) = 7.9 \times 10^{-42}\ L_{\rm H\alpha}\ ({\rm ergs\ s}^{-1}), \qquad (2)$$

where $L_{\rm [OII]}$ and $L_{\rm H\alpha}$ are the luminosities in [O II] $\lambda 3727$ and H$\alpha$ emission. We use the five band imaging from SDSS to estimate the photometric reddening, $E(B-V)_{(g-i)}$, for each galaxy as a whole. Using the Balmer decrement, we are also able to estimate the line-of-sight reddening, $E(B-V)_{H\alpha/H\beta}$. We correct for this reddening and also for the Milky Way extinction in each sight line. Table 1 lists the resultant SDSS SFR measurements.

---

[2] SPECP was developed by D. E. Welty and J. T. Lauroesch.

We note that the total SFR in the galaxy may be different from the above estimates, partly because the continuum of the galaxy is not known, since these lines are found superimposed on the quasar spectrum itself. However, the continuum contribution error is generally likely to be relatively low. The more significant reason the SFRs derived from the SDSS spectra may not be accurate is that the SDSS fiber itself typically probes only a fraction of the galaxy's projected surface area, except in the case of the very smallest galaxies. The low resolution of the SDSS images makes it difficult to precisely quantify what fraction of the galaxy surface area is covered by the SDSS fiber. We have, nevertheless made rough estimates of the SFR for the whole galaxy by scaling the dust-corrected SFRs estimated from the SDSS spectra by the ratio of the galaxy projected area as determined from the DECaLS images, and the projected area of the fiber, since in each of our fields, the angular separation of the quasar from the galaxy center is less than or nearly equal to the projected semi-major axis of the galaxy in arcseconds. The corrections thus estimated are substantial (more than a factor of 10) for four of our galaxies. These corrections are not precise, both because the DECaLS images are also modest in their spatial resolution, and because the SFR is unlikely to be uniform across the entire galaxy; however, the SFRs including these crude corrections are an improvement over the SFRs estimated purely from the SDSS spectra. In all of the statistical analysis presented in section 5, we use these "fiber loss-corrected" SFRs.

The distribution of stellar masses, estimated from the SDSS photometric data by Straka et al. (2015) from fitting stellar population synthesis models (see sec. 5.4.1 for more details) has a wide range (7.31≤log $M^*/M_\odot$≤9.78) with a mean stellar mass of log $M^*/M_\odot$=9.40 ± 0.36. However, we caution that these stellar mass estimates can be quite uncertain given the limited photometric information we have for these targets. Currently no near-infrared or UV photometric data are available and the SDSS images are quite shallow, and although we have performed PSF subtraction in all cases to remove the quasar contribution to the flux, the spatial resolution and depth of the SDSS images are not sufficient to obtain robust spectral energy distributions for our galaxies.

## 3. RESULTS

### 3.1. *Individual Objects*

#### 3.1.1. *Q0902+1414*

The COS spectrum for this quasar shows little, if any, flux. However, around 1500-1700 Å, there is a gradual drop off in the low-level flux present, indicating there is likely a partial Lyman Limit System (LLS) around $z_{LLS}$ ∼0.6 cutting off the flux below its Lyman limit. The GALEX FUV flux for this quasar is among the lowest in our sample (38 $\mu$Jy), and the redshift path length between the quasar and the galaxy is among the largest ($\Delta z$=0.9295), supporting the hypothesis that a higher-z LLS is blocking the quasar flux. Indeed, the SDSS spectrum of Q0902+1414 shows a potential system at $z$=0.633 with Mg I, Mg II, and Fe II absorption. We do no further analysis on the COS spectrum of this sightline. The DECaLS image of this galaxy shows an edge-on geometry with the quasar probing the north edge of the galaxy disk (Figure 2).

#### 3.1.2. *Q1005+5302*

We detect broad hydrogen absorption and metal lines associated with the foreground galaxy. Our H I column density measurements show a sub-DLA in this sightline with log $N_{HI}$ = $20.08^{+0.08}_{-0.07}$. Absorption features near the expected positions of the transitions for a number of low and intermediate metal ions (e.g., C I, C II, N I, O I, Al II, Si II, Si III, S II, Ni II, Co II) and high ions (C IV, Si IV) are also detected.

In an earlier work (Straka et al. 2015), we detected excess flux from the galaxy after quasar PSF subtraction, but were not able to determine the morphology or geometry of this galaxy due to its extremely small size and overlapping proximity to the quasar. Thus, the quasar appears to be directly probing the main body of the galaxy. See Figure 1.

### 3.1.3. *Q1010-0100*

The color of this quasar is extremely red, suggesting the presence of a large quantity of dust. The DLA along this sightline is flanked by geocoronal Ly$\alpha$ emission on the blue side and quasar O VI emission on the red side. We have left the quasar O VI emission in for the continuum fit of the spectrum. This is the strongest DLA in our sample with log $N_{HI}$=21.38±0.07. Absorption features near the expected positions of the transitions for several metal ions (e.g., C I, C II, O I, Mg I, Si II, Ni II) are also detected. Figure 1 suggests a face-on galaxy geometry with the quasar probing the outskirts of the visible disk.

### 3.1.4. *Q1130+6026*

We detect a DLA at the redshift of the foreground galaxy. There is contamination in the red wing of the DLA from geocoronal O I emission, as the absorption lies right against the emission line. This strongly affects the continuum fit for this system, and thus potentially causes an underestimation in the H I column density and errors in the fit compared to our other sightlines. With this in mind, we report a column density of log $N_{HI}$=20.57±0.04. Absorption features near the expected positions of the transitions for a number of low and intermediate metal ions (e.g., C II, N I, N V, O I, Al II, Si II, Ni II, Co II) are also detected. As seen in Figure 1, the foreground galaxy is too small to reliably determine morphology or geometry in this case.

### 3.1.5. *Q1135+2414*

The COS spectrum shows a DLA with log $N_{HI}$=20.57±0.03. Absorption features near the expected positions of the transitions for a number of low and intermediate metal ions (e.g., C I, C II, C IV, N I, O I, Al II, Si II, Si III, Si IV, S II, Ni II, Fe II) are also detected. As seen in Figure 1, this quasar probes a region ∼4 kpc from the center of a mostly face-on galaxy, on the outskirts of the star forming disk. According to the photometric properties derived from the SDSS r-band image, this galaxy is highly luminous with $L/L^*$=2.86.

### 3.1.6. *Q1328+2159*

For this target, the redshift of the galaxy was suitable to cover the wavelength ranges of both Ly$\alpha$ and Ly$\beta$ absorption. The spectrum shows a DLA in the blue wings of the background quasar's O VI emission line. We estimate log $N_{HI}$=21.01±0.11 from the Ly$\alpha$ and Ly$\beta$ absorption lines. Absorption features near the expected positions of the transitions for a number of low or intermediate metal ions (e.g., C I, C II, N I, O I, Al II, Si II, Si III, S II, Ni II, Fe II) as well as high ions (C IV, N V, Si IV) are also detected. Additionally, this system shows possible O VI absorption at the redshift of the galaxy. The small size of this galaxy seen in Figure 2 prevents us from reliably determining the geometry or morphology.

### 3.1.7. *Q1452+5443*

The redshift of this foreground galaxy allows for the use of Segment B. We detect a DLA with log $N_{HI} = 20.66^{+0.04}_{-0.03}$ cm$^{-2}$, constrained by the detection of Ly$\beta$. No O VI absorption is detected in Segment B, though we detect absorption features near the expected metal line positions for a number of low ions (e.g., C I, C II, N I, O I, Al II, Si II, Si III, S II, Ni II, Fe II, Co II) as well as high ions (C IV, Si IV). Additionally, there is an absorption line at $\lambda 1334$ Å that may be additional Ly$\alpha$ at more negative velocities, another interloping absorption system at a different redshift, or simply C II $\lambda 1334$ in the Milky Way interstellar medium. We exclude this absorption feature from our H I column density estimation.

This system is unique among our sample GOTOQs, as it has no detectable galaxy (in continuum light) in the SDSS images in any of the five filters (Figure 2). The upper limit on the stellar mass from photometric noise estimates outside the quasar PSF is $\log(M^*/M_\odot) < 8.0$. From the SDSS images, this quasar-galaxy configuration seems to be similar to two other pairs in this sample, Q1005+5302 and Q1130+6026. Both of the latter sight lines have compact galaxies that are difficult if not impossible to distinguish without proper quasar PSF subtraction. However, in the case of Q1452+5443, Straka et al. (2015) were unable to detect the foreground galaxy in the SDSS images even after PSF subtraction.

We therefore examined the DECaLS images of this field (deeper than the SDSS images), and found them to show a faint galaxy slightly offset from the quasar. We performed PSF subtraction on the r-band DECaLS image using a star in the same image for the PSF. The PSF-subtracted image shows an offset of $1.9''$ between the quasar and the brightest visible part of the galaxy, implying an impact parameter of 3.6 kpc at the redshift of the GOTOQ.

### 3.1.8. *Q1457+5321*

We detect no continuum flux from this quasar in the *HST* COS spectrum near the expected position of the Ly$\alpha$ absorption in the foreground galaxy at $z_{gal} = 0.0660$, and in fact a very low flux level below $\sim$1800 Å. The GALEX FUV flux for this quasar is among the lowest in our sample, (39 $\mu$Jy), and the redshift difference between the quasar and the galaxy is among the largest ($\delta z = 1.134$). The combination of these two factors likely indicates the presence of an LLS at $z\sim1$. Indeed, the SDSS spectrum of Q1457+5321 shows a potential system at $z = 0.99$ with Mg I, Mg II, and Fe II absorption. We do no further analysis on the COS spectrum of this sight line.

### 3.1.9. *Q1659+6202*

This galaxy appears to be roughly face-on, and has the highest stellar mass among our GOTOQs. We detect a DLA in this sightline with log $N_{HI}=21.20^{+0.07}_{-0.05}$. Absorption features near the expected positions of metal lines for many low ions (e.g., C I, C II, N I, O I, Si II, P II, S II, Ni II) as well as Si IV are detected for this system.

### 3.1.10. *Q2117+0026*

We detect a DLA in this sightline with log $N_{HI}=21.35^{+0.07}_{-0.06}$. There is some contamination from geocoronal O I emission in the red wing of the DLA. Absorption features near the expected positions of the metal lines for several low ions (e.g., C II, N I, O I, Si II, S II) as well as high ions (C IV, N V, Si IV) are detected for this system.

### 3.2. *Lyα Emission*

The SDSS fiber aperture and the COS aperture probe roughly the same region of the galaxy, being 3″ and 2.5″ in diameter respectively. Since we find strong nebular emission lines associated with the galaxy in this small region indicating active star formation, we might expect to find Lyα emission also. Indeed, previous works at 2<z<4 have found Lyα emission in the troughs of individual DLAs (e.g., Noterdaeme et al. 2012a; Kulkarni et al. 2012, 2015) and in stacked spectra of DLAs with log $N_{HI}$>21.0 (e.g. Noterdaeme et al. 2014; Joshi et al. 2017a). However, the DLA troughs for our absorbers do not show significant Lyα emission.

Table 3 lists the physical area probed by the COS aperture and the SDSS aperture at the redshift of the galaxy. It is possible that the Lyα emission from our galaxies is spatially offset from the regions covered by the COS aperture. We note in this context that, in the local universe, strong offsets between Lyα emission locations and stellar light locations have been observed (Kunth et al. 2003; Hayes et al. 2005; Östlin et al. 2009). Furthermore, Wisotzki et al. (2016) show that galaxies at 3<z<6 have diffuse Lyα emission halos with scale lengths of up to ∼7 kpc, and that between 40% and 90% of the total Lyα flux comes from the extended halo. Obtaining UV imaging of our galaxy sample may show that Lyα emission is extended and diffuse rather than concentrated within the instrument aperture due to the scattered nature of the emission line.

## 4. H I 21-CM EMISSION

Observations were carried out for 5 out of our 10 quasar sight lines with the Green Bank Telescope (GBT; PID AGBT16B_033, PI: N. Gupta), in order to determine the H I gas masses of the absorbing galaxies. The observations were carried out using the GBT's Gregorian focus L-band receiver and were spread over 8 observing runs during August - October, 2016 (see Table 4). The VErsatile GBT Astronomical Spectrometer (VEGAS) was used as the backend. A bandwidth of 11.720 MHz centered at the expected redshifted H I 21-cm line frequency was used. The band was split into 32,768 frequency channels. The data were acquired in two linear polarization channels, XX and YY, using a fast time sampling of 2 seconds. The observations were performed in standard position-switching mode with 2 minutes spent on-source and another 2 minutes off-source.

The data were processed based on NRAO's GBTIDL package following the procedures described in Gupta et al. (2012). First, the data were visually examined to identify and exclude scans with bad time stamps. After this, each scan was calibrated individually, and a first order baseline was fitted and subtracted to remove residual calibration and bandpass errors. The spectra from individual scans of the target source from all the observing runs were then combined to get the average XX and YY spectra. The XX and YY spectra were then combined to obtain the final stokes-I spectra. The corresponding spectral rms (1 $\sigma$) values are provided in column 5 of Table 4.

The H I 21-cm emission is detected in 2 cases, i.e., the $z = 0.0343$ galaxy toward Q1135+2414 and the $z = 0.0660$ galaxy toward Q1457+5321. In both cases, the broad H I 21-cm emission detections are somewhat redshifted with respect to the galaxy redshifts (the latter based on nebular emission lines), by ∼130-150 km s$^{-1}$. Fig. 5 shows these detections. The spectra presented in the figure have been smoothed to a spectral resolution of ∼20 km s$^{-1}$. The integrated line fluxes $\int S dv$ in these cases are 0.79 and 0.64 Jy km s$^{-1}$, respectively. Following the procedure outlined in Gupta et al. (2018), the corresponding total H I masses are estimated using the relation

$$M_{\rm HI} = 2.356 \times 10^5 \frac{D_L^2}{(1+z)} \int S dv. \tag{3}$$

The total H I masses thus estimated are summarized in column 6 of Table 4. For the three non-detections, 3 $\sigma$ upper limits on the H I mass, derived assuming an intrinsic emission line-width of 100 km s$^{-1}$, are listed.

Interestingly, the GOTOQs with H I 21-cm emission detections are in the sight lines to two of the top three most reddened background quasars in our sample of 10 GOTOQs [$E(B-V)$=0.36 and 0.39, respectively, for Q1135+2414 and Q1457+5321]. On the other hand, no 21-cm emission is detected at $z = 0.02$ in the field of Q1010-0100, while the H$\alpha$/H$\beta$ emission line ratio indicates a considerably larger amount of dust [$E(B-V)$=1.16] than the GOTOQs in the fields of Q1135+2414 or Q1457+5321; this indicates substantial variation in the 21-cm emission to $E(B-V)$ ratio. It will be interesting to obtain H I 21-cm emission measurements for a large sample of GOTOQs to examine whether such variations are common.

We also note that a weak anti-correlation has been noted between 21-cm absorption optical depth and the impact parameter in a study of galaxies in the fields of 21-cm absorbers at $z\sim0.3$ (Gupta et al. 2013; Dutta et al. 2017), including some GOTOQs. It will be interesting to study 21-cm absorption in a large sample of GOTOQs in order to understand how the properties of the cold neutral medium (CNM) in these galaxies correlate with the impact parameters, SFRs, and stellar masses of the galaxies, and with the absorber properties.

## 5. DISCUSSION

We now present the statistics of our H I detections, and discuss our sample in comparison with the literature, examining the correlations among the various galaxy and absorber properties.

### 5.1. *H I Column Densities and Velocity Offsets in our Sample*

The H I column densities for the absorbers in our sample are found to be in the range of log $N_{HI}$=20.08$^{+008}_{-0.07}$ to 21.38$\pm$0.07 in all 8 cases where the FUV flux of the quasar was detectable in the wavelength region near the Ly$\alpha$ absorption line at the GOTOQ redshift. Thus, our study has 100% success rate in detecting new DLA/sub-DLAs. The 50% fraction of DLAs with log $N_{HI}\geq$21.0 seems higher than may be expected based on the H I column density distribution of higher-redshift DLAs. But we caution that such comparisons may not be meaningful both because the GOTOQ sample is small, and because the selection strategies are very different for the GOTOQ sample (based on the galaxy) and the general DLA population (based on the absorber).

The velocity offsets of the absorption redshifts of the detected systems (as determined from the DLA fits) compared to the galaxy redshifts (determined from nebular emission lines) has a mean of -26.4 km s$^{-1}$ with a standard deviation of 106.5 km s$^{-1}$. [The large standard deviation reflects the fact that the velocity offset is much larger (ranging in absolute value from 88 to 185 km s$^{-1}$) for 4 of the 8 GOTOQs with detected DLA/sub-DLAs]. We note that given the low resolution of the COS G140L spectra ($\sim$200 km s$^{-1}$ at $\sim$1350 Å), it is not possible to determine the metal line velocity profiles accurately; therefore we cannot use the metal lines to determine the absorption redshifts. We note, however, that the absorption redshifts determined from the DLA profiles match closely with those determined from the Na I and Ca II absorption lines, in the few cases where these lines are detected in the SDSS spectra. We therefore adopt the redshifts estimated from the DLA fits, and caution that higher resolution spectra are needed to determine the metal absorption redshifts more accurately. We also note that the galaxy redshifts based on the nebular emission lines detected in the quasar spectra are measured away from the galaxy centers, at impact parameters ranging from 1.2 to 7.2 kpc. In any case, the offset values between the absorber redshifts from the DLA fits and the galaxy redshifts are well within the range of velocity spreads generally observed for DLA/sub-DLAs in higher-resolution spectroscopy, and there is little doubt that the detected absorbers are associated with the respective galaxies (or intervening cluster of galaxy-like objects at similar redshifts).

### 5.2. H I Column Densities and Reddening of the Background Quasars

It is also of interest to examine how the H I column densities and the dust in the absorbing galaxies correlate. One potential way to assess the dust content is via the reddening $E(B-V)$ of the background quasar. The ratio of the H I column density and reddening $N_{\rm HI}/E(B-V)$ for the GOTOQ toward Q1135+2414 is $1.03 \times 10^{21}$, far below the ratio of the total H column density to reddening $N_{\rm H}/E(B-V) = 5.8 \times 10^{21}$ observed in the Milky Way (Bohlin et al. 1978; Savage et al. 1977), possibly suggesting a predominance of $H_2$ in this sight line. In fact, only 2 of the 8 sight lines with $N_{\rm HI}$ measurements in our *HST* COS sample are consistent with even the nonlinear trend between $N_{\rm HI}/5.8 \times 10^{21}$ and $E(B-V)$ observed in the Milky Way ISM (e.g., Lizst 2014): four of the sight lines lie far above the Milky Way trend, and two others lie significantly below it. Part of the reason for the variation may be that the H I column densities (determined from Ly$\alpha$ absorption in the quasar sight line) trace the predominantly neutral gas, while the $E(B-V)$ values (estimated from the H$\alpha$/H$\beta$ emission line ratios) correspond to dust associated with the ionized gas in the star-forming regions, so a perfect correlation between the two quantities need not exist.

### 5.3. Comparison Sample from the Literature

We now combine our sample with the literature to search for trends between various galaxy and absorber properties, i.e., the impact parameter $\rho$, r-band luminosity relative to $L_*$, SFR, stellar mass, H I column density, absorber redshift, and velocity offset $\Delta v$ between the absorber and the galaxy, against each other (assuming the observed, intervening galaxies are the only ones contributing). For this purpose, we have compiled a comparison sample of spectroscopically confirmed galaxies detected at the same redshifts as DLA/sub-DLA absorbers based on the available measurements from the literature (Bowen et al. 2005; Christensen et al. 2005; Zwaan et al. 2005; Rosenberg et al. 2006; Peroux et al. 2011; Battisti et al. 2012; Noterdaeme et al. 2012a; Fynbo et al. 2013; Christensen et al. 2014; Som et al. 2015; Srianand et al. 2016; Péroux et al. 2016; Péroux et al. 2017; Zafar et al. 2017; Krogager et al. 2017; Neeleman et al. 2017; Augustin et al. 2018; Rhodin et al. 2018; Ma et al. 2018; Rahmani et al. 2018; Kanekar et al. 2018; Péroux et al. 2019; Frye et al. 2019; Borthakur et al. 2019; Mackenzie et al. 2019; Neeleman et al. 2019; Hamanowicz et al. 2020; Ranjan et al. 2020; Neeleman et al. 2020; Joshi et al. 2021; Rhodin et al. 2021; Duplis et al. 2021; Kaur et al. 2021, and references therein). The candidate galaxies for the $z$=0.0912 DLA in O I 363 and the $z$=0.0063 DLA in PG 1216+069 have been excluded because they have not been spectroscopically confirmed, and no impact parameter is known for them. The $z$=2.412 absorber toward Q0918+1636 was excluded because it has no continuum detection, and an emission-line detection of only [O III] $\lambda$5007, which is known to be a poor indicator of SFR (e.g., Kennicutt 1992; Moustakas et al. 2006). The full sample, including our GOTOQ sample, consists of 115 galaxies at redshifts $0 < z < 4.4$ believed to be associated with DLA/sub-DLA absorbers (either as the primary absorbing galaxies, or as members of the same group of galaxies).

We note that about 40% of the galaxies in this full sample (46 out of 115) are cases where multiple galaxies are detected at the redshift of the same DLA/sub-DLA (Peroux et al. 2011; Battisti et al. 2012; Christensen et al. 2005; Christensen et al. 2014; Augustin et al. 2018; Péroux et al. 2016; Péroux et al. 2019; Frye et al. 2019; Neeleman et al. 2019; Mackenzie et al. 2019; Borthakur et al. 2019; Hamanowicz et al. 2020; Kaur et al. 2021). In such cases, if the individual galaxies had measurements of luminosity, stellar mass, and/or SFR, we counted them individually. If only a combined luminosity, stellar mass, and/or SFR was known, the combined measurement was treated as a single object. This approach allows us to maximize the sample while searching for trends between the different stellar properties of the galaxies found in the fields of absorbers. (We note that most of the quantities presented in our examination of correlations in section 5.4 below are galaxy properties, not absorber properties.)

### 5.4. *Stellar Properties of the Galaxies*

#### 5.4.1. *Galaxies in Our Sample*

The mean impact parameter of the galaxy center from the quasar sightline for our GOTOQ sample is 4.1 kpc with a standard deviation of 1.8 kpc. All galaxies are within projected separations of $\leq 7.2$ kpc from the corresponding quasars. We note that the GOTOQ sample has small impact parameters by construction, since the galaxies were found by searching for emission lines in the SDSS fiber. We note that the 3″ diameter of the SDSS fibers corresponds to an impact parameter of 1.8 kpc at the median redshift of our sample (with a range of 0.65-3.6 kpc for our full sample).

The stellar masses of all the galaxies from our sample are adopted from Straka et al. (2015). These $M^*$ values are based on fitting of spectral energy distributions constructed using the available photometry in the five SDSS bands, using the photometric redshift code HYPERZ (Bolzonella, Miralles, & Pello 2000). HYPERZ uses the stellar population synthesis models of Bruzual & Charlot (2003) and minimizes the reduced $\chi^2$ to compute the best-fitting SED. The SFRs of our galaxies are based on measurements of the H$\alpha$ and [O II] $\lambda 3727$ emission lines (Straka et al. 2015), corrected by those authors for dust extinction using the H$\alpha$/H$\beta$ line ratio, and corrected for fiber losses using the approximate approach explained in sec. 2.2.2. The average SFR for our GOTOQ sample is 1.0 M$_\odot$ yr$^{-1}$ with a standard deviation of 1.4 M$_\odot$ yr$^{-1}$. Followup high-resolution multi-band imaging or integral field spectroscopy is necessary for us to understand the global stellar population and star formation history of these galaxies.

#### 5.4.2. *Galaxies in the Comparison Sample*

For a substantial fraction of the galaxies from the comparison sample, the stellar masses are direct measurements based on spectral energy distribution (SED) fitting of the galaxies. In the remaining cases where the stellar mass based on SED-fitting was not available, if the SFR was available, $M^*$ was estimated from the SFR based on the redshift evolution of the star-forming main sequence, as parameterized by Boogaard et al. (2018) for $0<z<0.5$[3] and by Whitaker et al. (2014, 2020) for $0.5<z<2.5$[4], and adopting the latter for $z>2.5$ as well. If neither SED-fitting results nor the SFR were available, $M^*$ was estimated from the luminosity. To do this, we used an empirical calibration

---

[3] log SFR = 0.83 log $(M^*/10^{8.5} M_\odot)$ -0.83 + 1.74 log $([1+z]/1.55)$

[4] log SFR = a + b log $(M^*/M_\odot)$ +c [log $(M^*/M_\odot)]^2$, with the higher-precision values of the redshift-dependent coefficients $a$, $b$, and $c$ tabulated in (Whitaker et al. 2020)

$\log (M^*/M_\odot) = 1.13 \log (L/L_*) + 9.85$, based on a linear regression between $\log (L/L_*)$ and $\log M^*$ for all 23 galaxies in our sample (including all of our GOTOQs and 13 absorbing galaxies from the literature) for which independent estimates of the stellar mass and the luminosity were available. (See sec. 5.5 and Fig. 7 below for more discussion of how well the stated relations fit the observed data for DLA/sub-DLA galaxies.)

In total, we have stellar mass estimates for ∼86% of our full sample. We consider the values from SED fitting (available for ∼ 54% of our sample) as more reliable, and regard the rest as indicative. While searching for trends between the various parameters, we present statistics for both, the subset with stellar mass values based on SED fitting ($M^*_{\rm SED}$) and the overall set including the stellar mass values based on SFR or luminosity ($M^*_{\rm adopt}$).

### 5.5. *Correlations among galaxy and absorber properties*

We now combine our GOTOQ sample with the literature sample in an attempt to search for trends between the various galaxy and absorber properties, since both of these samples are ultimately cases of galaxies and absorbers associated with each other. Of course, we note that these two samples were selected in different ways, with the GOTOQ sample being galaxy-selected, and the literature sample being primarily absorber-selected. Furthermore, due to the the requirement to have detectable emission lines in the SDSS fiber, the GOTOQ sample selects galaxies that are star-forming, at least to some extent; the literature sample has no such constraint. A more direct comparison using uniform selection criteria and covering the whole multi-dimensional parameter space (in redshift, impact parameter, stellar mass, luminosity, SFR, etc.) would of course be ideal, but the currently available samples do not allow that. Furthermore, we emphasize that we are not looking at how populated the different regions of the parameter space are, but merely at whether there are any trends between the various parameters. We also note that even within the literature sample, different studies differ in measurement techniques (e.g., narrow-band imaging, long-slit spectroscopy, or integral field spectroscopy) and in the rest-frame wavelength regions used for observing the absorber galaxies (e.g., optical or far-infrared/sub-mm). Indeed, in a few of the cases where both rest-frame UV and far-IR data exist, the SFRs based on far-IR (e.g., the continuum near [C II] 158 $\mu$m emission) observations have been found to be higher than those based on the near-UV continuum, indicating the presence of some dust extinction, though in general dust extinction seems to be relatively modest (e.g., Kaur et al. 2021).

Fig. 6 shows the H I column densities and absorption redshifts of the DLAs/sub-DLAs detected in this study and those at $z_{abs}$<0.35 from the literature. The Spearman rank order correlation coefficient is $r_S$ = -0.132 with a high two-tailed probability $P$=0.528 that this value of $r_S$ arises purely by chance. Thus, no significant correlation is observed between $N_{\rm HI}$ and $z_{abs}$ at $z_{abs}$<0.35.

The top panel of Fig. 7 shows the stellar mass based on SED fitting vs. SFR for all 43 galaxies in our sample for which these measurements are available (including all of our GOTOQs and 33 other galaxies associated with DLA/sub-DLAs from the literature mentioned in sec. 5.3). For comparison the SFMS of Boogard et al. (2018) and Whitaker et al. (2014) are also shown. A substantial fraction of DLA /sub-DLA galaxies appear to be generally consistent with the SFMS trends for galaxies. Indeed, this also justifies our use of these relations for estimating the stellar masses of some other galaxies for which SED fitting was not available but SFR measurements were available. The bottom panel of Fig. 7 shows the stellar mass based on SED fitting vs. luminosity data for all 23 galaxies in our sample for which these measurements are available (including all of our GOTOQs and 13 other galaxies associated with DLA/sub-DLAs from the literature). Also shown is our best fit based on linear regression (mentioned above in sec. 5.4.2). Overall, this appears to be a good fit to the available data, justifying our use of this relation to estimate stellar masses for some other galaxies where neither SED-fitting results nor the SFR were available.

Fig. 8 shows plots of various properties of the full sample to examine correlations. The cyan circles denote the GOTOQs from this work, while the orange squares denote the measurements from the literature. Table 5 shows the results of correlation tests between the various parameters of interest, listing the Spearman rank order correlation coefficient $r_S$ between pairs of parameters, and the corresponding probability of the observed value of $r_S$ occurring purely by chance. A number of statistically significant correlations (with a $<0.05$ probability of occurring purely by chance, i.e. without the existence of an underlying correlation) are evident in many of the panels of Fig. 8. A positive correlation is observed between the stellar mass and impact parameter (as seen in the second panel from left in the fourth row from the top in Fig. 8). This may potentially indicate that more massive galaxies can have gas-rich regions detectable as DLA/sub-DLAs out to larger galactocentric distances.

A positive correlation is also seen between luminosity and stellar mass (third panel from left in the fourth row from top in Fig. 8), and between SFR and stellar mass (fourth panel from left in the fourth row from top in Fig. 8). These latter relations arise partly because of the assumption of the SFMS or the observed stellar mass vs. luminosity relation for estimating $M^*$ values in some cases. We note that the $M^*$ values in Fig. 8 refer to the adopted stellar masses described earlier (including values derived from SED fitting or other techniques). However, as seen from Table 5, the trends between luminosity and $M^*$ or between SFR and $M^*$ are not altered much if only the $M^*$ values derived directly from SED fitting are included. Overall, these positive correlations suggest that a substantial fraction of DLA /sub-DLA galaxies are generally consistent with the stellar mass-luminosity relation and the SFMS of star-forming galaxies. We return to a discussion of the SFMS in section 5.5.5 below.

### 5.5.1. *Trends in H I Column Density*

We now consider trends between various galaxy properties and the H I column density. In cases where multiple galaxies are detected at the redshift of the same DLA/sub-DLA, in principle, the contribution of each galaxy to the H I column density should be used. However, individual velocity components cannot be discerned in the Ly$\alpha$ absorption profiles (which are dominated by the broad damping wings). Higher Lyman series lines are not always covered, and even if they are, they can be blended with other Ly forest lines, making it difficult to determine the velocity structure of the H I. It is thus not possible to determine H I column densities associated with the individual galaxies in cases where multiple galaxies are present at nearly the same redshift as the DLA/sub-DLA. Therefore, we use the same $N_{\rm HI}$ of the absorber for each galaxy associated with the absorber. (This is why there are some "horizontal lines" parallel to the $N_{\rm HI}$ axis in most of the panels in the lowest row of Fig. 8 (which show log $N_{\rm HI}$ plotted vs. the various other quantities.) We note that this is unavoidable with the existing data, and that this approach has also been adopted in other works in the literature (e.g., Hamanowicz et al. 2020).

The H I column density is negatively correlated with the impact parameter (as seen in the second panel from left in the bottom row of Fig. 8). This trend has been seen in smaller samples before and for other types of absorbers (such as Mg II absorbers, e.g., Nielsen et al. 2013, Huang 2021). Our study is the first one to examine this trend in such a large sample, and a sample focusing specifically on DLA/sub-DLAs. We also emphasize that our sample includes many absorbers at very low impact parameters. The decrease in H I column density with increasing impact parameter probably results from a decreasing probability of encountering high column density gas at increasing galactocentric radii. Our sample shows that the anti-correlation between the H I column density and impact parameter appears to extend even to the smallest impact parameters probed ($\lesssim 1$ kpc). Given the scatter in the data, it will be very interesting to increase the $N_{\rm HI}$ measurements at low-impact parameters further to begin examining the structure of the cool ISM in the inner regions of galaxies. Such measurements will also offer statistical probes of the radial gradients in H I column density and eventually in metallicity, and allow an examination of redshift evolution of these gradients.

We also note in this context that a weak anti-correlation has been noted between 21-cm absorption optical depth and the impact parameter in a study of galaxies in the fields of 21-cm absorbers at $z\sim 0.3$ (Gupta et al. 2013; Dutta et al. 2017), including some GOTOQs. It would be interesting to study 21-cm absorption in a larger sample of GOTOQs in order to understand how the properties of the cold neutral medium (CNM) in these galaxies correlate with the SFR and stellar mass of the galaxy.

Interestingly, as seen from Table 5, there is also a tentative negative correlation between the H I column density and the stellar mass determined from SED fitting.[5] We note that a similar negative trend between the H I column density and the stellar mass was previously suggested in a much smaller sample (Augustin et al. 2018). We also note in this context the suggestion in previous works that sub-DLAs may arise in higher-mass galaxies than DLAs (e.g., Khare et al. 2007; Kulkarni et al. 2010). The negative correlation we observe between $N_{\rm HI}$ and $M^*_{SED}$ is consistent with this suggestion. We emphasize that this trend includes the full redshift range $0<z<4.4$, and is not limited to the low-$z$ sub-DLAs whose high metallicities were partly the reason behind the suggestion of Khare et al. (2007) and Kulkarni et al. (2010).

### 5.5.2. *Observed Dependence on Redshift*

In Fig. 9, we examine the observed trends in various properties with absorber redshift. There is a large "hole" in redshift space, which highlights the sparse coverage of absorbers in the range $1 \lesssim z \lesssim 1.8$. This sparse coverage highlights the difficulties in identifying galaxies associated with DLA/sub-DLAs at both low and high redshifts. At $z<1.65$, the Ly$\alpha$ line (needed for determining the H I column density) lies in the UV, making it necessary to use *HST* to discover these DLA/sub-DLAs. The practical necessity to target relatively bright quasars for UV spectroscopy with *HST* limits the number of DLA/sub-DLAs observed at $z<1.65$. Furthermore, of the DLA/sub-DLAs detected in this redshift range, those at $z \lesssim 1$ are more easily identified with galaxies, since the latter are easier to image, making it easier to determine their stellar properties.

At $1.65 \lesssim z \lesssim 4.5$, the determination of H I column density is easier since the DLA/sub-DLA Ly$\alpha$ lines are redshifted into the optical range accessible to ground-based spectrographs. On the other hand, the imaging and spectroscopic confirmation of the absorbing galaxies is much harder due to the rapidly decreasing surface brightness of galaxies with redshift, and due to the difficulty of spatially resolving the background quasar from the absorbing galaxy, which again limits the sample of DLA/sub-DLA galaxies. While great strides are being made in the imaging of galaxies associated with DLA/sub-DLAs using integral field spectroscopy (IFS), e.g. with VLT MUSE and SINFONI (e.g., Peroux et al. 2011; Péroux et al. 2016; Péroux et al. 2019; Mackenzie et al. 2019), the redshift ranges of the observed galaxies are limited by the wavelength coverage of the IFS for detecting nebular emission lines of the galaxies. The systems at $z>3.8$ are based on ALMA detections of [C II] 158 $\mu$m or CO (2-1) emission (e.g., Neeleman et al. 2017, 2019, 2020). We nevertheless attempt to understand the dependence on redshift of the absorbing galaxies using the data at higher and lower redshift, while emphasizing the need to increase the samples at all redshifts, and especially to fill the hole in the "redshift desert", given the importance of the epoch $1<z<2$ in the cosmic star formation history.

---

[5] A similar trend is not seen with $M^*_{adopt}$, i.e., if galaxies with stellar masses determined from other techniques are included. Even if the $M^*_{adopt}$ values for multiple galaxies corresponding to a given absorber are added together and treated as one single galaxy, the combined $M^*_{adopt}$ values do not show a correlation with $N_{\rm HI}$. However, we note that the other stellar mass estimates included in $M^*_{adopt}$ are less reliable than those based on SED fitting. Indeed, most of these other stellar mass estimates are based on SFR, which does not show much correlation with the H I column density. In any case, given that the stellar masses based on SED fitting are more reliable, we regard the negative correlation between $N_{\rm HI}$ and $M^*_{SED}$ as likely to be real.

The first and third panels from the top in Fig. 9, showing galaxy luminosities and stellar masses, respectively, do not exhibit strong correlations in these quantities with respect to redshift. This reality underscores the fact that the absorption-line technique used to identify most of the literature sample is independent of the galaxy luminosity and mass.

Interesting trends are, however, seen in some of the remaining panels of Fig. 9, despite the considerable scatter in the data. The second panel from the top shows the SFR of the galaxy plotted vs. the redshift. A trend of SFR increasing with redshift is observed over the redshift range covered. This suggests that the galaxies associated with DLAs and sub-DLAs (and those located in the same groups as the main absorbing galaxies) have star formation histories (SFHs) consistent with the SFH of star-forming galaxies at least at $z \lesssim 1$. The relatively flat SFR vs. $z$ trend at $z \gtrsim 2$ appears a little surprising given that the global SFR density inferred from emission-based studies of star-forming galaxies shows a well-defined peak at $z \sim 2$ and a considerable decrease between $z \sim 2$ and $z \sim 4.5$ (e.g., Madau & Dickinson 2014). Increasing the samples of galaxies associated with DLA/sub-DLAs at $z>2$ will of course help to more accurately establish the redshift dependence trend at these redshifts. We discuss the star formation history in more detail in section 5.5.4 below.

The bottom panel in Fig. 9 shows the H I column density plotted vs. redshift for the full sample. The dashed black line in this panel denotes the separation between DLAs and sub-DLAs. Most of the sub-DLAs in the sample are at $z<1.1$, while the DLAs cover the full range of redshifts. This may not be a physical effect, but a result of the fact that a much smaller number of sub-DLAs have been studied even in absorption at $z>2$, and only a small minority of these have detections of the associated galaxies. This effect is probably responsible for the significant positive correlation observed between the H I column density and redshift in the absorbers associated with galaxies. (We note that the correlation between H I column density and redshift is evaluated only for "non-duplicate" objects, i.e., treating multiple galaxies detected at the same absorber redshift as one since, as explained in sec. 5.5.1, only the total H I column density is known and assigned to all of them.)

### 5.5.3. *Gas Kinematics*

The fourth panel from the top in Fig. 9 shows the velocity offset $\Delta v$ of the absorber relative to the galaxy vs. the absorber redshift. For our GOTOQs, we use the centers of the H I Ly$\alpha$ profiles to determine the absorber redshifts. This is because the metal line redshifts are not precisely known for the GOTOQs because of the very low resolution of the COS G140L grating that prevents accurate measurements of the metal line profiles. The dashed black line in this panel separates the cases of absorbers at higher velocities with respect to the galaxies from those at lower velocities. No significant correlation is observed between $\Delta v$ and redshift. 3/4 of the absorbers (6 out 8) in our GOTOQ sample appear to be blueshifted ($\Delta v<0$) with respect to the galaxies. However, the overall sample (including the comparison sample) shows a roughly equal incidence of positive and negative velocity offsets.

We caution that non-zero velocity offsets could arise partly from the fact that our absorption line measurements are based on the H I Ly$\alpha$ profiles measured in the COS G140L spectra. But given the low resolution and low S/N of these COS spectra, we do not consider it advisable to use the metal line wavelengths as an indicator of the exact absorber redshift. We also note that the absorption redshifts we adopt from the H I Ly$\alpha$ lines agree well with those deduced from Ca II or Na I absorption lines in cases where the latter are detected in SDSS spectra. But high-resolution measurements of the metal lines are clearly essential for more accurate determinations of the absorber redshifts. If a dominance of absorbers with $\Delta v<0$ persists in our GOTOQ sample even with high-resolution spectroscopy, and is confirmed in a larger GOTOQ sample, it may provide further insights into the kinematics of the absorbing gas in the GOTOQs.

Absorption at $\Delta v<0$ can occur in outflowing gas from the near side of a galaxy, or from inflowing gas on the far side of the galaxy, or from gas co-rotating with the disk of a galaxy. The galaxy inclination and the orientation of the quasar sight line relative to the galaxy major axis also matter in this context since, for example, gas outflows from galaxies are often believed to occur primarily along the minor axis, and gas inflows primarily along the major axis. Thus, negative velocity shifts could arise in outflows from the near side of a face-on galaxy or inflows on the far side of an edge-on galaxy, or in gas in the approaching side co-rotating with the disk of an edge-on galaxy. By contrast, positive velocity shifts can arise in outflows from the far side of a face-on galaxy or inflows on the near side of an edge-on galaxy, or in gas in the receding side co-rotating with the disk of an edge-on galaxy. Indeed, complex gas motions have been detected in some of the galaxies in our sample based on integral field spectroscopy (e.g., Peroux et al. 2011; Péroux et al. 2019). It will be interesting to sort out among the different possibilities for each individual case by combining the results from high-resolution spectroscopy of the background quasar with integral field spectroscopy of the foreground galaxy.

### 5.5.4. *Star Formation History*

Fig. 10 examines in more detail the star formation history for the galaxies in our full sample for which stellar mass estimates based on SED fitting results are available. The upper panel shows the specific star formation rate (sSFR = SFR/$M^*_{SED}$) for the galaxies for which SED fitting results are available, plotted vs. redshift. While there is some scatter in the data points, there is an overall positive trend between sSFR and $\log(1+z)$, with a Spearman rank order correlation coefficient of 0.452 and a probability of $2.33\times10^{-3}$ of this value of $r_S$ occurring purely by chance. While the sample is small, the existing data suggest the existence of multiple sSFR vs. redshift sequences, for galaxies of different stellar masses. This is consistent with the multiple sSFR vs. redshift trends reported by Whitaker et al. (2014) for galaxies of different stellar masses. The dashed magenta and green curves show the sSFR vs. redshift trends of Whitaker et al. (2014) for log $M^*$=9.3 and 11.1, respectively. The range in sSFR spanned by DLA/sub-DLA galaxies is larger. The existence of some DLA/sub-DLA galaxies with sSFRs above the curve for log $M^*$=9.3 is a result of some of the galaxies being lower in mass (as low as log $M^*$=7.1). However, it is surprising that some DLA/sub-DLA galaxies (∼20% of the galaxies plotted in Fig. 10) lie below the curve for even log $M^*$=11.1, with the difference being substantial for ∼15%. Given that the stellar masses of these galaxies are in fact lower than log $M^*$=11.1, their low sSFRs suggest that their star formation activity is substantially below the expectations based on the SFMS. Some of these galaxies have stellar masses 10<log $M^*$<11 and moderately low SFRs, while some others have stellar masses 9<log $M^*$<10 and very low SFRs. In either case, these DLA/sub-DLAs are associated with galaxies with longer gas consumption time scales than the usual star-forming galaxies.

##### 5.5.5. *Star Formation Main Sequence*

The lower panel of Fig. 10 shows the specific star formation rate vs. the stellar mass for the galaxies in our full sample for which stellar mass estimates based on SED fitting results are available. Once again, while there is considerable scatter, overall there is a significant negative correlation. The Spearman rank order correlation coefficient is -0.548 with a probability of $1.42 \times 10^{-4}$ of this occurring purely by chance. The large scatter appears to result from different SFR vs. stellar mass relations at different redshifts. The dashed red and blue curves show the SFMS for galaxies at 0.5<$z$<1.0 and 2.0<$z$<2.5 from Whitaker et al. (2014). The DLA/sub-DLA galaxies span a much wider range in sSFR and stellar mass than those covered in Whitaker et al. (2014), extending to stellar masses as low as log $M^*$=7.1. Furthermore, the DLA/sub-DLA sample covers a larger redshift range (0<$z$<4.4) than the range 0.5<$z$<2.5 covered by Whitaker et al. (2014). For comparison, we also show the sSFR vs. stellar mass relations determined by Thorne et al. (2021) that do extend down to lower masses and cover a wider redshift range. The green and magenta curves show the SFMS of Thorne et al. (2021) for the redshift ranges 0.02<$z$<0.08 and 4.25<$z$<5.0. These relations nicely bracket the range of sSFR covered by most DLA/sub-DLA galaxies. However, there are a few DLA/sub-DLA galaxies that have sSFRs below the range of the Thorne et al. (2021) relations. This may indicate the existence of some galaxies with lower star formation activity than expected from the SFMS for typical star-forming galaxies.

Thus, both panels of Fig. 10 illustrate the power of the quasar absorption line technique to extend galaxy evolution studies to low sSFRs and low stellar masses. Future studies of larger samples of DLA/sub-DLAs will help to grow this population that may be missed by studies of emission-selected galaxies.

##### 5.5.6. *Gas Mass vs. Stellar Mass*

Fig. 11 compares the H I gas mass and stellar mass for DLA/sub-DLA galaxies and local galaxies. The upper panel shows the H I mass vs. stellar mass for the galaxies associated with DLAs/sub-DLAs from our study and the literature (Kanekar et al. 2018; Borthakur et al. 2019) for which stellar mass measurements exist, and H I 21-cm emission was searched for. Together, these consist of 8 21-cm detections and 4 upper limits in the redshift range of 0.01$\lesssim z \lesssim$0.10. The detections (large blue circles) show a strong positive trend, with a Spearman rank order correlation coefficient of 0.952 and a probability of $2.60 \times 10^{-4}$ of this value of $r_S$ occurring purely by chance.

The remaining data points in Fig. 11 show measurements for local (0<$z$<0.03) galaxies from the literature. These include galaxies from the THINGS survey (Leroy et al. 2008), the ALLSMOG survey (Bothwell et al. 2014), and a sample of gas-rich galaxies (McGaugh 2012). The DLA/sub-DLA data are seen to be overall consistent with the trends for local galaxies, with H I mass and stellar mass following the trend expected from the standard scaling relation. The dashed line shows the trend fitted by Calette et al. (2018) for late-type galaxies (LTGs, including spirals and irregulars). The DLA/sub-DLA galaxies lie above the LTG trend, extending to much lower stellar masses than the local spirals do, but fall in the general range of stellar masses for the gas-rich galaxies. While there is considerable scatter in the relation for both the DLA/sub-DLA galaxies and the local galaxies, most (6 of the 8) DLA/sub-DLA galaxies with 21-cm detections show a very tight relation, and lie at the upper end of the trends for the local galaxies, even the gas-rich local galaxies. In other words, most DLA/sub-DLA galaxies with 21-cm detections appear to have higher H I masses compared to the typical local galaxies with comparable stellar masses. This could potentially be an artifact of the small sample size; expanding the sample is thus essential to verify this trend.

The lower panel of Fig. 11 shows the ratio of the H I mass to stellar mass for the DLA/sub-DLA galaxies and the comparison with values for local galaxies. The H I to stellar mass ratio falls off with increasing stellar mass for DLA/sub-DLA galaxies, as it does for local galaxies. However, the ratio is much higher for DLA/sub-DLA galaxies than for typical local spirals and irregulars. In fact, even at the low stellar masses, the H I mass to stellar mass ratio is higher for DLA/sub-DLA galaxies than for most of the gas-rich galaxies. This is consistent with our suggestion that DLA/sub-DLAs trace galaxies with longer gas depletion time scales than most star-forming galaxies. It is essential to obtain H I 21-cm emission observations for a much larger DLA and sub-DLA sample to examine whether the trends seen in the DLA/sub-DLA sample presented here also hold for the overall DLA/sub-DLA population. It would also be very interesting to determine the gas mass to stellar mass ratio for higher redshift DLA/sub-DLA galaxies and examine how this ratio has evolved with cosmic time.

The dotted lines in both panels of Fig. 11 show the trend fitted by Calette et al. (2018) for early-type galaxies (ETGs). The DLA/sub-DLA galaxies lie far above the ETG trend, suggesting that most of the DLA galaxies with H I 21-cm detections (and possibly even those with 21-cm limits) are not ETGs. Of course, the existing DECaLS images of our GOTOQs (e.g., Fig. 1) show that not all of our GOTOQs are obviously late-type galaxies. However, these images are relatively shallow and low in spatial resolution. It is essential to obtain high-S/N and high-resolution images of the DLA/sub-DLA galaxies to determine the morphology more robustly and to assess whether the H I mass to stellar mass trends in these galaxies are consistent with the trends for local galaxies of different morphological types.

Fig. 12 compares the H I gas masses and stellar masses of DLA/sub-DLA galaxies with nearby H I-selected galaxies. The upper panel shows the H I gas masses plotted vs. stellar masses for the DLA/sub-DLA galaxies, along with the corresponding relation observed for galaxies selected by means of their H I emission, based on a sample of 9,153 H I-selected ALFALFA galaxies with SDSS spectra and stellar masses from the MPA-JHU catalog (Maddox et al. 2015). The H I masses for most DLA/sub-DLAs with 21-cm detections (6 out of 8) are higher than would be expected for their stellar masses based on the trend observed for the spectroscopic ALFALFA-SDSS sample, and higher than even the upper envelope of the latter trend. Such a trend could potentially result from a selection effect if the DLA/sub-DLA galaxies were somehow chosen to be more luminous than the ALFALFA galaxies, e.g., if the DLA/sub-DLA galaxies were systematically at higher redshifts than the ALFALFA galaxies. However, the redshift range $z<0.08$ and the median redshift of 0.026 for the 8 DLA/sub-DLA galaxies with H I masses and stellar masses shown in Fig. 12 are roughly similar to those for the ALFALFA surey (which has a redshift range of $z<0.06$ and a median redshift of $\sim0.03$). Thus, observational selection cannot explain the higher brightness (in H I 21 cm emission) of the DLA/sub-DLA galaxies compared to the ALFALFA galaxies apparent from the upper panel of Fig. 12.

Another possibility is that the detected 21-cm emission signal includes other bright galaxies in the field of view, given that the full width at half maximum (FWHM) of the GBT beam at 1.4 GHz is approximately 9.3′ in diameter. To assess this possibility, we searched for galaxies within a 9.3′ diameter region of the two GOTOQs with 21-cm emission detections, with similar g magnitude as our GOTOQ or brighter than that. For Q1457+531, there are two such galaxies: an early-type galaxy with no SDSS spectrum (but a photometric redshift of 0.135) 2.77′ away from the quasar and a late-type galaxy at $z=0.0264$ separated by 4.05′ from the quasar. Given the large redshift differences in both cases from the redshift of our GOTOQ ($z_{gal}=0.066$), both are unlikely to contribute much to the H I 21-cm emission detected with GBT. For Q1135+2414 ($z_{gal}=0.0343$), there is a galaxy with no SDSS spectrum but a photometric redshift of 0.154 located 4.9′ away from the quasar. This galaxy also cannot contribute much to the GBT signal, both because its photometric redshift differs substantially from the redshift of our GOTOQ, and because it lies outside the FWHM of the GBT field of view. Thus, we conclude that our GBT 21-cm emission detections are associated with the corresponding GOTOQs, and that the corresponding estimates of H I masses of these galaxies should be robust. This suggests that low-$z$ DLA/sub-DLA galaxies with H I and stellar mass determinations are intrinsically more gas-rich than typical local galaxies. Of course, given the small size of the current sample, it is essential to obtain H I 21-cm emission observations for many more DLA/sub-DLAs to examine whether the trends observed here also hold for the overall DLA/sub-DLA population.

The lower panel of Fig. 12 shows a plot of the ratio of H I mass to stellar mass vs. the stellar mass for the DLA/sub-DLA galaxies and the corresponding trend for the spectroscopic ALFALFA-SDSS galaxies. The $M_{\rm HI}/M^*$ ratio flattens substantially at stellar masses below $\sim 10^9$ $M_\odot$ for the spectroscopic ALFALFA-SDSS sample. On the other hand, no such flattening is evident for DLA/sub-DLA galaxies. Larger samples of DLA/sub-DLA galaxies (including those with high stellar masses) are essential to determine whether the $M_{\rm HI}/M^*$ ratio shows a flattening at any stellar mass for these galaxies. The flattening in $M_{\rm HI}/M^*$ observed for the ALFALFA-SDSS galaxies at $M^* \sim 10^9$ $M_\odot$ is believed to be connected to the fact that this stellar mass marks the transition between the hot and cold-mode accretion in simulations at a baryonic mass of $\sim 2\text{-}3\times 10^{10}$ $M_\odot$ (e.g., Keres et al. 2009). The ALFALFA-SDSS galaxies with $M^* < 10^9$ $M_\odot$ are metal-poor and generally irregular in morphology, suggesting that their high H I fractions originate in recent gas accretion rather than inefficient star formation (Maddox et al. 2015). The even higher H I fractions for DLA/sub-DLA galaxies, together with their high sSFRs seen at $M^* < 10^9$ $M_\odot$ (see Fig. 10), suggest that DLA/sub-DLA galaxies may also have undergone recent gas accretion. It will be very interesting to obtain higher-resolution UV spectra in the future to determine the metallicities of these DLA/sub-DLAs and assess whether they are consistent with the recent accretion scenario (low metallicities would be expected if the gas is recently accreted).

We also note that the logarithmic H I to stellar mass ratios for the Leo P and Leo T dwarf galaxies are ≈0.25 and ≈0.4, respectively. The flattening of the $M_{\rm HI}/M^*$ ratio at low stellar masses has been suggested as an evidence that dark galaxies (which are rich in H I but low in stellar mass) are not a substantial population (Maddox et al. 2015). The lack of flattening in $M_{\rm HI}/M^*$ seen for DLA/sub-DLAs, if confirmed with a larger sample, would raise questions about this interpretation, especially if the rise in $M_{\rm HI}/M^*$ continues to even lower stellar masses. Expanding optical and radio studies of DLAs/sub-DLAs may thus have important implications for dark galaxies.

## 6. CONCLUSIONS

We have targeted ten quasars selected from our GOTOQs sample with HST/COS in order to study the H I characteristics of the galaxy disks. Eight of these sight lines had sufficient quasar flux at the wavelengths of interest, and all 8 of these showed strong H I absorption: 7 DLAs and one sub-DLA . Thus, we have a 100% detection rate of DLA or sub-DLA systems in the disks of low−$z$ galaxies. Indeed, this is the largest systematically targeted sample of DLAs associated with low-redshift galaxies. Additionally, our low-resolution COS spectra provide approximate measurements of metal absorption lines and their equivalent widths. A search for H I 21-cm emission measurements for five of the fields led to two detections and three non-detections.

Combining our sample with the literature, we obtained a sample of 115 galaxies associated with DLAs/sub-DLAs and examined the correlations between different galaxy and absorber properties. While this includes both galaxy-selected and absorber-selected samples, the combined sample allows us to examine basic scaling relations between galaxy and absorber properties, and reveals a number of interesting trends:

(1) The SFR correlates positively with redshift overall, with a strong rise at lower redshifts and a possible flattening at $z \gtrsim 2$. It would be interesting to expand the high-$z$ sample to determine the star formation history of DLA/sub-DLA galaxies more accurately.

(2) The H I column density is inversely correlated with the impact parameter, consistent with past findings based on smaller samples. Our sample shows that this trend appears to extend even to the smallest impact parameters probed ($\lesssim 1$ kpc). This confirms the general expectation that the probability of a random sightline to a background quasar passing through a region gas-rich enough to be detectable as a DLA/sub-DLA decreases with increasing galactocentric distance.

(3) The H I column density is inversely correlated with the stellar mass, in agreement with past findings based on smaller samples. This trend is also consistent with past suggestions that sub-DLAs may arise in more massive galaxies than DLAs.

(4) Roughly equal incidences of positive and negative velocity offsets are observed between the absorbers and the galaxies. A combination of high-resolution spectroscopy and integral field spectroscopy is needed to determine the origins of the absorbing gas in individual cases.

(5) A positive correlation is observed between the stellar mass and the impact parameter, possibly indicating that more massive galaxies can have gas-rich regions detectable as DLA/sub-DLAs out to larger galactocentric distances.

(6) A substantial fraction of DLA/sub-DLA galaxies with independent estimates of stellar mass and SFR and/or stellar mass and luminosity appear to be generally consistent with the SFR vs. $M^*$ and luminosity vs. $M^*$ trends for star-forming galaxies.

(7) The sSFR shows a positive correlation with redshift and a negative correlation with stellar mass. However, while the sSFRs for most DLA/sub-DLA galaxies are broadly consistent with the expectations from the SFMS for star-forming galaxies, a significant fraction of the DLA/sub-DLA galaxies (∼20%) lie below the SFMS. This illustrates how studies of DLA/sub-DLAs can complement studies of galaxy evolution and extend them to lower stellar masses and lower SFRs.

(8) Indeed, the relevance of DLA/sub-DLAs for the study of gas-rich galaxies is also evident from the higher H I mass to stellar mass ratio for most DLA/sub-DLA galaxies with detections of H I 21-cm emission compared to local spiral and irregular galaxies, and even relative to the H I emission-selected ALFALFA-SDSS galaxies. The high $M_{\rm HI}/M^*$ ratio and high sSFRs seen in DLA/sub-DLA galaxies with $M^* < 10^9 M_\odot$, suggest that these galaxies may be gas-rich because of recent gas accretion rather than inefficient star formation.

To make further progress on the nature of the gas flows around these galaxies, it is imperative to determine the metallicity of the absorbing gas. Follow-up higher-resolution and shorter-wavelength UV spectra are essential to more accurately measure the column densities of the metal lines, and to cover higher Lyman series lines. Absorption-line metallicities obtained from future higher-resolution *HST* spectroscopic studies can be compared with metallicities of the galaxies based on nebular emission lines to determine the metallicity gradients in the galaxies. It is also essential to estimate H I masses of galaxies using radio interferometers to overcome the potential for confusion that can affect single dish H I 21-cm measurements. The ongoing large H I 21-cm emission line surveys such as MALS (Gupta et al. 2016) and WALLABY (Koribalski20) with the Square Kilometre Array (SKA) precursor telescopes will deliver such samples of nearby galaxies to understand the origin of gas producing DLA/subDLAs.

The metallicities and dust contents of GOTOQs inferred from future studies can be combined with the remaining properties discussed in this paper for determination of the mass-metallicity relation (MZR) of the low-$z$ DLA/sub-DLAs, and to assess how the MZR has been evolving with redshift for the absorbing galaxies. The metallicities will also help to better understand whether the DLA/sub-DLAs with H I 21-cm detections indeed experienced recent accretion of metal-poor gas. Rest-frame UV and optical continuum imaging of more DLA/sub-DLAs is essential to increase the sample with luminosity measurements that can be combined with the stellar mass measurements to investigate the mass-to-light ratios of the absorbing galaxies. Also necessary are optical spectra of the galaxies to determine their total flux profile, and high resolution *HST* imaging to ascertain the exact morphological profiles of these galaxies. Followup high-resolution spectroscopy of the background quasars, combined with integral field spectroscopy of the galaxies, would also be instrumental in determining the kinematics of the cool gas and warm ionized gas in these galaxies.

Finally, it is also important to fill in the large gaps in redshift space seen in Figs. 9 and 10a, especially at $1 \lesssim z \lesssim 2$. This along with an increase in the existing number of galaxies associated with DLAs and sub-DLAs at $z<1$ and $z>2$ is essential to build a more comprehensive picture of the evolution of the DLA and sub-DLA populations and to better understand their overall significance in studies of galaxy evolution, for example for better understanding the physics involved in the transformation of the cold gas into stars in galaxies. Our work demonstrates that absorption studies of DLA/sub-DLAs combined with emission studies of the galaxies producing them have the power to probe regimes of low stellar mass and low SFR that are not sampled well with current flux-limited galaxy surveys.

## ACKNOWLEDGMENTS

This work is based on observations with the NASA/ESA *Hubble Space Telescope* obtained at the Space Telescope Science Institute (STScI), which is operated by the Association of Universities for Research in Astronomy, Incorporated, under NASA contract NAS5-26555, and the Green Bank Observatory, which is a facility of the National Science Foundation operated under cooperative agreement by Associated Universities, Inc.. Support for *HST* program GO-14137 (PI: Straka) was provided through a grant from the STScI under NASA contract NAS5-26555. VPK acknowledges additional partial support from the National Science Foundation grants AST/2007538 and AST/2009811 (PI: Kulkarni). LAS acknowledges support from the European Research Council Grant agreement 278594-GasAroundGalaxies. DGY acknowledges partial support from the John Templeton Foundation grants 12711 and 37426 (PI: York). This research has made use of NASA's Astrophysics Data System. This research has also made use of the NASA/IPAC Extragalactic Database (NED) which is operated by the Jet Propulsion Laboratory, California Institute of Technology, under contract with NASA.

*Facilities:* HST(COS), GBT, SDSS

Table 1. Quasar-Galaxy Pairs Observed with HST COS

| Quasar | RA, Dec (J2000) | $z_{quasar}$ | $z_{gal}$ | $f^1_{Ly\alpha}$ (cgs) | $m_{rgal}$ | $L/L^{*2}_r$ | $\log M^{*3}$ ($M_\odot$) | $\theta^4$ (") | $\rho^5$ (kpc) | $SFR^6_{SDSS}$ ($M_\odot$ yr$^{-1}$) | Ref, ID$^7$ |
|---|---|---|---|---|---|---|---|---|---|---|---|
| Q0902+1414 | 09:02:50.47, 14:14:08.29 | 0.980 | 0.0505 | <0.1 | 17.41 | 1.21 | 9.65 | 3.6 | 3.6 | 0.03 | S15, 4 |
| Q1005+5302 | 10:05:14.21, 53:02:40.04 | 0.560 | 0.1358 | 7.0 | 19.50 | 0.31 | 9.60 | 1.5 | 3.6 | 0.19 | Y12, 12 |
| Q1010-0100 | 10:10:15.74, -01:00:38.11 | 0.230 | 0.0213 | 1.5 | 17.91 | 0.03 | 8.88 | 2.9 | 1.2 | 0.04 | Y12, 2 |
| Q1130+6026 | 11:30:02.99, 60:26:28.62 | 0.370 | 0.0604 | 4.0 | 20.28 | 0.02 | 7.31 | 1.7 | 2.0 | 0.02 | S13, 15 |
| Q1135+2414 | 11:35:55.66, 24:14:38.10 | 1.450 | 0.0343 | 4.0 | 16.48 | 2.86 | 9.41 | 5.7 | 3.9 | 0.01 | S15, 18 |
| Q1328+2159 | 13:28:24.33, 21:59:19.66 | 0.330 | 0.1352 | 2.5 | 19.12 | 0.25 | 9.06 | 2.4 | 5.7 | 0.16 | S15, 31 |
| Q1452+5443 | 14:52:40.53, 54:43:45.11 | 1.520 | 0.1026 | 4.0 | … | … | 7.96 | 1.9 | 3.6 | 0.25 | Y12, 18 |
| Q1457+5321 | 14:57:19.00, 53:21:59.27 | 1.200 | 0.0660 | <0.1 | 17.95 | 0.74 | 9.28 | 3.4 | 4.2 | 0.04 | S15, 40 |
| Q1659+6202 | 16:59:58.94, 62:02:18.14 | 0.230 | 0.1103 | 13 | 18.53 | 0.48 | 9.78 | 3.6 | 7.2 | 0.26 | S13, 24 |
| Q2117-0026 | 21:17:01.31, -00:26:38.80 | 1.137 | 0.0580 | 1.9 | 17.32 | 0.38 | 9.64 | 5.1 | 5.7 | 0.04 | S13, 25 |

1. Continuum flux of the background quasar in units of $10^{-16}$ erg s$^{-1}$ cm$^{-2}$ Å$^{-1}$ near the wavelength of the Ly-α absorption from the foreground galaxy, as measured in our HST COS spectra. 2. The r-band L/L* values of our foreground galaxies found in SDSS. Values <0.2 are considered dwarf galaxies. No value is listed for Q1452+5443 since the SDSS images do not resolve the foreground galaxy. 3. Stellar mass of the galaxy, adopted from Straka et al. (2015). 4. Angular separation of the galaxy from the sight line of the quasar in arcseconds, determined from the SDSS images, except for Q1452+5443 where the value measured from DECaLS images is listed. 5. Impact parameter of the galaxy with the sight line of the quasar in kpc determined from the SDSS images, except for Q1452+5443 where the value measured from DECaLS images is listed. 6.Dust-corrected SFR determined from Hα emission line flux detected in the SDSS fibers, from Straka et al. (2015). 7. Reference for galaxy magnitude and the ID number for the galaxy in that reference. Y12: York et al. (2012), S13: Straka et al. (2013); S15: Straka et al. (2015).

Table 2. Metal-line Measurements for the DLA/Sub-DLAs in the COS spectra

| Quasar | $\lambda_{obs}$ (Å) | $W^1_{obs}$ (mÅ) | $\sigma W^{phot,2}_{obs}$ (mÅ) | $\sigma W^{cont,3}_{obs}$ (mÅ) | $\sigma W^{tot,4}_{obs}$ (mÅ) | $W^5/\sigma W_{tot}$ | Ion | $\lambda_{rest}$ |
|---|---|---|---|---|---|---|---|---|
| Q1005+5302 | 1108.892 | 1788.65 | 292.86 | 171.44 | 339.35 | 5.3 | O I | 976.448 |
| | 1123.251 | 1523.63 | 186.05 | 170.24 | 252.19 | 6.0 | O I | 988.773 |
| | 1176.995 | 2867.97 | 265.16 | 261.89 | 372.68 | 7.7 | C II | 1036.336 |
| | 1287.801 | 289.07 | 53.00 | 43.95 | 68.85 | 4.2 | N I | 1134.980 |
| | 1351.561 | 740.60 | 37.90 | 35.39 | 51.85 | 14.3 | Si II | 1190.415 |
| | 1354.938 | 734.41 | 40.11 | 42.39 | 58.36 | 12.6 | Si II | 1193.289 |
| | 1362.602 | 468.06 | 37.89 | 43.30 | 57.54 | 8.1 | N I | 1199.549 |
| | 1369.652 | 1159.75 | 44.84 | 45.97 | 64.22 | 18.1 | Si III | 1206.500 |
| | 1420.093 | 392.88 | 32.38 | 37.67 | 49.68 | 7.9 | S II | 1250.578 |
| | 1423.493 | 264.16 | 28.81 | 31.02 | 42.33 | 6.2 | S II | 1253.805 |
| | 142 9.803 | 255.39b | 24.73 | 21.31 | 32.65 | 7.8 | S II | 1259.518 |
| | 1431.017 | 1028.38b | 31.01 | 17.62 | 35.66 | 28.8 | Si II | 1260.422 |
| | 1478.377 | 711.24 | 52.72 | 56.65 | 77.39 | 9.2 | O I | 1302.168 |
| | 1480.637 | 732.69 | 58.95 | 76.79 | 96.80 | 7.6 | Si II | 1304.370 |
| | 1495.201 | 177.41 | 31.51 | 23.16 | 39.10 | 4.5 | Ni II | 1317.217 |
| | 1508.578 | 187.61 | 38.59 | 37.74 | 53.98 | 3.5 | C I | 1328.833 |
| | 1514.666 | 1629.26b | 61.42 | 51.52 | 80.16 | 20.3 | C II | 1334.532 |
| | 1516.344 | 352.85b | 42.27 | 40.41 | 58.48 | 6.0 | C II* | 1335.707 |
| | 1581.801 | 644.88 | 56.75 | 50.09 | 75.69 | 8.5 | Si IV | 1393.760 |
| | 1592.546 | 226.20 | 48.40 | 47.70 | 67.96 | 3.3 | Si IV | 1402.772 |
| | 1645.636 | 415.14b? | 73.48 | 72.76 | 103.41 | 4.0 | Co II | 1448.019 |

Table 2 *(continued)*

| Quasar | $\lambda_{obs}$ (Å) | $W^1_{obs}$ (mÅ) | $\sigma W^{phot,2}_{obs}$ (mÅ) | $\sigma W^{cont,3}_{obs}$ (mÅ) | $\sigma W^{tot,4}_{obs}$ (mÅ) | $W^5/\sigma W_{tot}$ | Ion | $\lambda_{rest}$ |
|---|---|---|---|---|---|---|---|---|
|  | 1650.646 | 295.69 | 65.73 | 60.52 | 89.35 | 3.3 | Ni II | 1454.842 |
|  | 1667.570 | 275.18 | 67.75 | 66.75 | 95.11 | 2.9 | Ni II | 1467.756 |
|  | 1680.841 | 309.37 | 61.46 | 43.05 | 75.04 | 4.1 | Co II | 1480.954 |
|  | 1733.266 | 666.14 | 88.71 | 88.23 | 125.12 | 5.3 | Si II | 1526.707 |
|  | 1757.432 | 603.73 | 83.41 | 61.07 | 103.37 | 5.8 | C IV | 1548.204 |
|  | 1760.094 | 127.16 | 62.17 | 49.69 | 79.59 | 1.6 | C IV | 1550.781 |
|  | 1771.365 | 341.38 | 76.14 | 63.13 | 98.91 | 3.5 | C I | 1560.309 |
|  | 1881.290 | 269.29 | 76.65 | 61.40 | 98.21 | 2.7 | C I | 1656.928 |
|  | 1896.862 | 461.32 | 85.03 | 64.92 | 106.98 | 4.3 | Al II | 1670.788 |
| Q1010−0100 | 1286.944 | 544.46 | 96.07 | 89.10 | 131.03 | 4.2 | Si II | 1260.422 |
|  | 1329.792 | 597.68 | 112.32 | 71.56 | 133.18 | 4.5 | O I | 1302.168 |
|  | 1333.477 | 362.52 | 105.07 | 80.01 | 132.06 | 2.7 | Si II | 1304.370 |
|  | 1362.575 | 617.21 | 114.44 | 77.05 | 137.96 | 4.5 | C II | 1334.532 |
|  | 1367.030 | 461.28b? | 100.06 | 60.07 | 116.71 | 4.0 | C II* | 1335.707 |
|  | 1593.518 | 1539.74 | 230.43 | 210.75 | 312.27 | 4.9 | C I | 1560.309 |
|  | 1777.170 | 1407.64 | 326.32 | 227.88 | 398.01 | 3.5 | Ni II | 1741.553 |
|  | 1785.543 | 501.86 | 178.75 | 48.95 | 185.33 | 2.7 | Mg I | 1747.793 |
| Q1130+6026 | 1262.454 | 209.25 | 41.04 | 33.26 | 52.83 | 4.0 | Si II | 1190.415 |
|  | 1265.428 | 570.90 | 63.80 | 77.42 | 100.32 | 5.7 | Si II | 1193.289 |
|  | 1272.433 | 275.15 | 50.26 | 51.64 | 72.06 | 3.8 | N I | 1199.549 |
|  | 1273.901 | 259.41b? | 45.89 | 41.67 | 61.99 | 4.2 | N I | 1200.223 |
|  | 1314.466 | 408.24 | 65.87 | 73.58 | 98.76 | 4.1 | N V | 1238.821 |

**Table 2** (continued)

| Quasar | $\lambda_{obs}$ (Å) | $W_{obs}^1$ (mÅ) | $\sigma W_{obs}^{phot,2}$ (mÅ) | $\sigma W_{obs}^{cont,3}$ (mÅ) | $\sigma W_{obs}^{tot,4}$ (mÅ) | $W^5/\sigma W_{tot}$ | Ion | $\lambda_{rest}$ |
|---|---|---|---|---|---|---|---|---|
| | 1317.113 | 136.34 | 45.56 | 38.29 | 59.51 | 2.3 | N V | 1242.804 |
| | 1336.447 | 429.06 | 54.94 | 43.57 | 70.12 | 6.1 | Si II | 1260.422 |
| | 1380.934 | 517.94 | 55.67 | 52.64 | 76.61 | 6.8 | O I | 1302.168 |
| | 1383.399 | 735.70 | 63.12 | 64.44 | 90.20 | 8.2 | Si II | 1304.370 |
| | 1415.262 | 303.29 | 54.33 | 58.96 | 80.17 | 3.8 | C II | 1334.532 |
| | 1553.652 | 402.93 | 82.57 | 67.77 | 106.82 | 3.8 | Co II | 1466.211 |
| | 1555.467 | 394.59 | 79.95 | 62.20 | 101.29 | 3.9 | Ni II | 1467.259 |
| | 1570.084 | 379.03 | 65.27 | 32.27 | 72.82 | 5.2 | Co II | 1480.954 |
| | 1653.712 | 448.95 | 107.33 | 125.06 | 164.81 | 2.7 | C I | 1560.309 |
| | 1756.726 | 584.00 | 130.23 | 69.86 | 147.79 | 4.0 | C I | 1656.928 |
| | 1813.409 | 726.43 | 134.82 | 61.14 | 148.03 | 4.9 | Ni II | 1709.604 |
| | 1890.756 | 961.66 | 261.48 | 128.27 | 291.25 | 3.3 | P I | 1782.829 |
| | 1896.507 | 683.70 | 189.70 | 24.39 | 191.26 | 3.6 | P I | 1787.648 |
| Q1135+2414 | 1173.254 | 837.39 | 85.85 | 109.14 | 138.85 | 6.0 | N I | 1134.980 |
| | 1184.169 | 425.50 | 75.71 | 97.25 | 123.24 | 3.5 | Fe II | 1144.937 |
| | 1195.672 | 257.11 | 62.36 | 67.72 | 92.06 | 2.8 | C I | 1155.809 |
| | 1231.278 | 494.57 | 53.35 | 61.97 | 81.77 | 6.0 | Si II | 1190.415 |
| | 1234.175 | 498.62 | 52.32 | 58.66 | 78.60 | 6.3 | Si II | 1193.289 |
| | 1240.691 | 497.20 | 40.52 | 27.28 | 48.85 | 10.2 | N I | 1199.549 |
| | 1241.627 | 401.57b | 36.25 | 21.67 | 42.23 | 9.5 | N I | 1200.223 |
| | 1242.362 | 355.06 | 36.25 | 23.53 | 43.22 | 8.2 | N I | 1200.709 |
| | 1248.032 | 589.87 | 46.79 | 39.29 | 61.10 | 9.7 | Si III | 1206.500 |
| | 1293.760 | 261.32b? | 48.26 | 55.93 | 73.87 | 3.5 | S II | 1250.578 |

Table 2 (continued)

| Quasar | $\lambda_{obs}$ (Å) | $W_{obs}^1$ (mÅ) | $\sigma W_{obs}^{phot,2}$ (mÅ) | $\sigma W_{obs}^{cont,3}$ (mÅ) | $\sigma W_{obs}^{tot,4}$ (mÅ) | $W^5/\sigma W_{tot}$ | Ion | $\lambda_{rest}$ |
|---|---|---|---|---|---|---|---|---|
| | 1296.946 | 942.40b? | 68.25 | 94.98 | 116.96 | 8.1 | S II | 1253.805 |
| | 1303.423 | 468.56b | 58.64 | 79.24 | 98.58 | 4.8 | Si II | 1260.422 |
| | 1347.071 | 591.12 | 48.71 | 36.17 | 60.67 | 9.7 | O I | 1302.168 |
| | 1348.950 | 794.46 | 60.75 | 61.75 | 86.62 | 9.2 | Si II | 1304.370b |
| | 1380.658 | 681.96 | 59.91 | 53.81 | 80.53 | 8.5 | C II | 1334.532 |
| | 1441.647 | 335.00 | 53.02 | 38.67 | 65.63 | 5.1 | Si IV | 1393.760 |
| | 1449.950 | 280.40b? | 56.96 | 50.60 | 76.19 | 3.7 | Si IV | 1402.772 |
| | 1462.950 | 179.48 | 53.02 | 47.43 | 71.14 | 2.5 | Ga II | 1414.402 |
| | 1579.564 | 1738.11 | 156.91 | 187.03 | 244.14 | 7.1 | Si II | 1526.707 |
| | 1591.126 | 1228.62b? | 129.40 | 123.20 | 178.67 | 6.9 | Si II* | 1533.431 |
| | 1601.577 | 691.27 | 104.08 | 88.56 | 136.66 | 5.1 | C IV | 1548.204 |
| | 1603.635 | 389.36 | 91.49 | 80.79 | 122.05 | 3.2 | C IV | 1550.781 |
| | 1612.704 | 342.32 | 76.91 | 46.60 | 89.93 | 3.8 | C I | 1560.309 |
| | 1663.800 | 633.27 | 95.61 | 62.72 | 114.35 | 5.5 | Fe II | 1608.451 |
| | 1714.113 | 142.93 | 61.34 | 36.33 | 71.29 | 2.0 | C I | 1656.928 |
| | 1728.560 | 800.06 | 108.76 | 82.90 | 136.76 | 5.9 | Al II | 1670.788 |
| | 1737.626 | 654.82 | 101.08 | 74.83 | 125.76 | 5.2 | P I | 1679.696 |
| | 1761.507 | 369.94 | 99.14 | 57.19 | 114.45 | 3.2 | Ni II | 1703.411 |
| Q1328+2159 | 1123.523 | 2031.42b | 588.96 | 406.42 | 715.57 | 2.8 | Si II | 989.873 |
| | 1171.543 | 763.93 | 223.87 | 77.64 | 236.96 | 3.2 | O VI | 1031.926 |
| | 1177.093 | 2983.72 | 438.33 | 290.52 | 525.86 | 5.7 | C II | 1036.336 |
| | 1287.867 | 751.99 | 116.43 | 117.85 | 165.66 | 4.5 | N I | 1134.414 |
| | 1299.963 | 261.83 | 65.31 | 80.54 | 103.69 | 2.5 | Fe II | 1144.937 |

Table 2 continued on next page

Table 2 (continued)

| Quasar | $\lambda_{obs}$ (Å) | $W_{obs}^1$ (mÅ) | $\sigma W_{obs}^{phot,2}$ (mÅ) | $\sigma W_{obs}^{cont,3}$ (mÅ) | $\sigma W_{obs}^{tot,4}$ (mÅ) | $W^5/\sigma W_{tot}$ | Ion | $\lambda_{rest}$ |
|---|---|---|---|---|---|---|---|---|
| | 1308.604 | 81.56 | 13.56 | 10.01 | 16.85 | 4.8 | P II | 1152.818 |
| | 1312.193 | 398.92 | 73.99 | 93.87 | 119.52 | 3.3 | C I | 1155.809 |
| | 1351.215 | 774.32 | 91.98 | 137.17 | 165.15 | 4.7 | Si II | 1190.415 |
| | 1354.667 | 793.15 | 75.11 | 76.36 | 107.10 | 7.4 | Si II | 1193.289 |
| | 1362.543 | 1143.60 | 66.37 | 84.34 | 107.32 | 10.7 | N I | 1199.549 |
| | 1369.706 | 1049.62 | 57.51 | 56.84 | 80.86 | 13.0 | Si III | 1206.500 |
| | 1407.496 | 425.55 | 91.66 | 117.80 | 149.26 | 2.9 | N V | 1238.821 |
| | 1410.345 | 337.65 | 69.29 | 61.02 | 92.33 | 3.7 | N V | 1242.804 |
| | 1423.638 | 196.75 | 56.11 | 43.27 | 70.86 | 2.8 | S II | 1253.805 |
| | 1430.594 | 1363.88 | 105.66 | 110.07 | 152.58 | 8.9 | Si II | 1260.422 |
| | 1478.275 | 596.06 | 96.05 | 96.98 | 136.49 | 4.4 | O I | 1302.168 |
| | 1481.245 | 474.28b | 78.43 | 58.85 | 98.06 | 4.8 | Si II | 1304.370 |
| | 1515.329 | 1572.56 | 160.66 | 185.82 | 245.64 | 6.4 | C II | 1334.532 |
| | 1582.422 | 1072.72b? | 93.86 | 129.06 | 159.58 | 6.7 | Si IV | 1393.760 |
| | 1603.268 | 148.44 | 34.83 | 30.53 | 46.32 | 3.2 | Ni II | 1412.866 |
| | 1605.737 | 105.30 | 28.43 | 20.10 | 34.82 | 3.0 | Ga II | 1414.402 |
| | 1757.356 | 1363.27 | 158.70 | 73.00 | 174.69 | 7.8 | C IV | 1548.204 |
| | 1759.953 | 1018.83 | 161.85 | 125.54 | 204.83 | 5.0 | C IV | 1550.781 |
| | 1825.907 | 1047.00 | 158.57 | 52.65 | 167.08 | 6.3 | Fe II | 1608.451 |
| | 1896.112 | 1640.99 | 291.16 | 163.73 | 334.04 | 4.9 | Al II | 1670.788 |
| Q1452+5443 | 1171.726 | 690.91b? | 189.44 | 128.35 | 228.83 | 3.0 | Fe II | 1063.176 |
| | 1312.090 | 247.47 | 40.60 | 35.38 | 53.86 | 4.6 | Si II | 1190.415 |
| | 1315.236 | 577.85b? | 49.70 | 44.42 | 66.66 | 8.7 | Si II | 1193.289 |

Table 2 *(continued)*

| Quasar | $\lambda_{obs}$ (Å) | $W^1_{obs}$ (mÅ) | $\sigma W^{phot,2}_{obs}$ (mÅ) | $\sigma W^{cont,3}_{obs}$ (mÅ) | $\sigma W^{tot,4}_{obs}$ (mÅ) | $W^5/\sigma W_{tot}$ | Ion | $\lambda_{rest}$ |
|---|---|---|---|---|---|---|---|---|
| | 1321.609 | 367.87 | 39.10 | 27.07 | 47.55 | 7.7 | N I | 1199.549 |
| | 1322.751 | 722.81 | 47.30 | 31.88 | 57.04 | 12.7 | N I | 1200.223 |
| | 1323.596 | 127.75 | 28.70 | 17.51 | 33.62 | 3.8 | N I | 1200.709 |
| | 1329.819 | 577.04 | 69.92 | 52.79 | 87.62 | 6.6 | Si III | 1206.500 |
| | 1379.656 | 131.18 | 36.57 | 26.86 | 45.38 | 2.9 | S II | 1250.578 |
| | 1381.584 | 407.20 | 59.73 | 68.92 | 91.20 | 4.5 | S II | 1253.805 |
| | 1389.292 | 1661.66b | 88.73 | 116.52 | 146.46 | 11.3 | Si II | 1260.422 |
| | 1394.919 | 204.98 | 43.83 | 40.08 | 59.39 | 3.5 | Si II* | 1264.737 |
| | 1435.556 | 898.99 | 58.77 | 55.16 | 80.60 | 11.2 | O I | 1302.168 |
| | 1437.779 | 1229.33 | 66.64 | 67.80 | 95.06 | 12.9 | Si II | 1304.370 |
| | 1471.121 | 1838.09b? | 89.63 | 89.61 | 126.74 | 14.5 | C II | 1334.532 |
| | 1483.332 | 565.69n | 79.24 | 111.92 | 137.13 | 4.1 | Ni II | 1345.878 |
| | 1536.460 | 328.43 | 56.89 | 47.79 | 74.30 | 4.4 | Si IV | 1393.760 |
| | 1683.181 | 612.38 | 98.75 | 71.69 | 122.03 | 5.0 | Si II | 1526.707 |
| | 1706.528 | 720.49 | 87.82 | 55.00 | 103.62 | 7.0 | C IV | 1548.204 |
| | 1709.616 | 987.15b | 103.52 | 77.54 | 129.34 | 7.6 | C IV | 1550.781 |
| | 1736.325 | 476.75 | 85.38 | 68.26 | 109.31 | 4.4 | Co II | 1574.550 |
| | 1841.708 | 604.02 | 145.91 | 108.05 | 181.56 | 3.3 | Al II | 1670.788 |
| | 1885.903 | 587.57 | 133.69 | 43.12 | 140.47 | 4.2 | Ni II | 1709.604 |
| | 1919.265 | 370.32 | 113.36 | 57.81 | 127.25 | 2.9 | Ni II | 1741.553 |
| | 1931.124 | 1154.98b? | 171.78 | 90.21 | 194.03 | 6.0 | Ni II | 1751.915 |
| Q1659+6202 | 1279.329 | 319.26 | 69.22 | 74.22 | 101.48 | 3.1 | P II | 1152.818 |
| | 1286.556 | 161.43 | 39.97 | 40.11 | 56.63 | 2.9 | C I | 1157.909 |

Table 2 (continued)

| Quasar | $\lambda_{obs}$ (Å) | $W_{obs}^1$ (mÅ) | $\sigma W_{obs}^{phot,2}$ (mÅ) | $\sigma W_{obs}^{cont,3}$ (mÅ) | $\sigma W_{obs}^{tot,4}$ (mÅ) | $W^5/\sigma W_{tot}$ | Ion | $\lambda_{rest}$ |
|---|---|---|---|---|---|---|---|---|
| | 1321.172 | 307.83b? | 24.86 | 14.33 | 28.70 | 10.7 | S III | 1190.203 |
| | 1321.740 | 346.17b? | 28.30 | 20.23 | 34.79 | 10.0 | Si II | 1190.415 |
| | 1324.864 | 735.41 | 40.82 | 41.56 | 58.25 | 12.6 | Si II | 1193.289 |
| | 1333.011 | 1793.01b | 55.63 | 58.78 | 80.93 | 22.2 | N I | 1199.549 |
| | 1388.469 | 226.62 | 30.68 | 27.37 | 41.12 | 5.5 | S II | 1250.578 |
| | 1391.981 | 225.09 | 31.78 | 29.89 | 43.62 | 5.2 | S II | 1253.805 |
| | 1398.352 | 194.98b? | 25.94 | 18.52 | 31.87 | 6.1 | S II | 1259.518 |
| | 1399.428 | 1040.39b? | 44.21 | 38.76 | 58.80 | 17.7 | Si II | 1260.422 |
| | 1445.627 | 881.06 | 37.19 | 40.96 | 55.33 | 15.9 | O I | 1302.168 |
| | 1448.062 | 639.13 | 35.41 | 41.14 | 54.28 | 11.8 | Si II | 1304.370 |
| | 1481.908 | 1500.84 | 33.78 | 37.98 | 50.83 | 29.5 | C II | 1334.532 |
| | 1463.251 | 620.76 | 37.42 | 55.90 | 67.27 | 9.2 | Ni II | 1317.217 |
| | 1495.086 | 209.77 | 18.51 | 20.80 | 27.85 | 7.5 | Ni II | 1345.878 |
| | 1521.052 | 198.26 | 24.26 | 26.54 | 35.96 | 5.5 | Ni II | 1370.132 |
| | 1547.135 | 591.20b? | 35.98 | 31.72 | 47.97 | 12.3 | Si IV | 1393.760 |
| | 1557.247 | 307.65 | 34.22 | 35.53 | 49.33 | 6.2 | Si IV | 1402.772 |
| Q2117-0026 | 1146.336 | 550.65 | 110.70 | 69.39 | 130.65 | 4.2 | N II | 1083.993 |
| | 1262.129 | 1064.35b | 104.44 | 94.70 | 140.98 | 7.5 | Si II | 1193.289 |
| | 1269.541 | 1133.87b | 112.84 | 118.35 | 163.52 | 6.9 | N I | 1200.223 |
| | 1310.159 | 138.09 | 54.34 | 51.00 | 74.52 | 1.9 | N V | 1238.821 |
| | 1315.415 | 433.45b | 62.11 | 52.07 | 81.05 | 5.3 | N V | 1242.804 |
| | 1323.359 | 505.95b | 85.84 | 89.36 | 123.92 | 4.1 | S II | 1250.578 |
| | 1326.531 | 717.25b | 85.84 | 74.87 | 113.90 | 6.3 | S II | 1253.805 |

**Table 2** *(continued)*

| Quasar | $\lambda_{obs}$ (Å) | $W_{obs}^1$ (mÅ) | $\sigma W_{obs}^{phot,2}$ (mÅ) | $\sigma W_{obs}^{cont,3}$ (mÅ) | $\sigma W_{obs}^{tot,4}$ (mÅ) | $W^5/\sigma W_{tot}$ | Ion | $\lambda_{rest}$ |
|---|---|---|---|---|---|---|---|---|
| | 1332.281 | 378.27 | 66.96 | 49.58 | 83.31 | 4.5 | S II | 1259.518 |
| | 1333.463 | 663.46 | 71.57 | 40.79 | 82.38 | 8.1 | Si II | 1260.422 |
| | 1377.631 | 401.40 | 89.00 | 81.67 | 120.79 | 3.3 | O I | 1302.168 |
| | 1380.075 | 348.15 | 69.62 | 41.62 | 81.11 | 4.3 | Si II | 1304.370 |
| | 1412.108 | 1152.08 | 126.92 | 131.81 | 182.98 | 6.3 | C II | 1334.532 |
| | 1411.955 | 958.22 | 107.27 | 83.01 | 135.64 | 7.1 | C II | 1334.532 |
| | 1474.881 | 807.27 | 113.97 | 110.91 | 159.03 | 5.1 | Si IV | 1393.760 |
| | 1483.744 | 582.64 | 116.09 | 136.46 | 179.16 | 3.3 | Si IV | 1402.772 |
| | 1615.124 | 924.61b | 125.58 | 83.16 | 150.62 | 6.1 | Si II | 1526.707 |
| | 1637.733 | 1084.52 | 154.45 | 162.22 | 223.99 | 4.8 | C IV | 1548.204 |
| | 1639.857 | 654.51 | 129.79 | 125.83 | 180.77 | 3.6 | C IV | 1550.781 |

1. Observed-frame equivalent width in mÅ. A "b" or "b?" after the equivalent width value denotes a blend or a possible blend with another unrelated absorption feature. 2. 1 $\sigma$ uncertainty in the measured equivalent width in mÅ due to photon noise. 3. 1 $\sigma$ uncertainty in the measured equivalent width in mÅ due to uncertainty in continuum determination. 4. Combined 1-$\sigma$ uncertainty in the measured equivalent width in mÅ. 5 Significance level of the absorption feature.

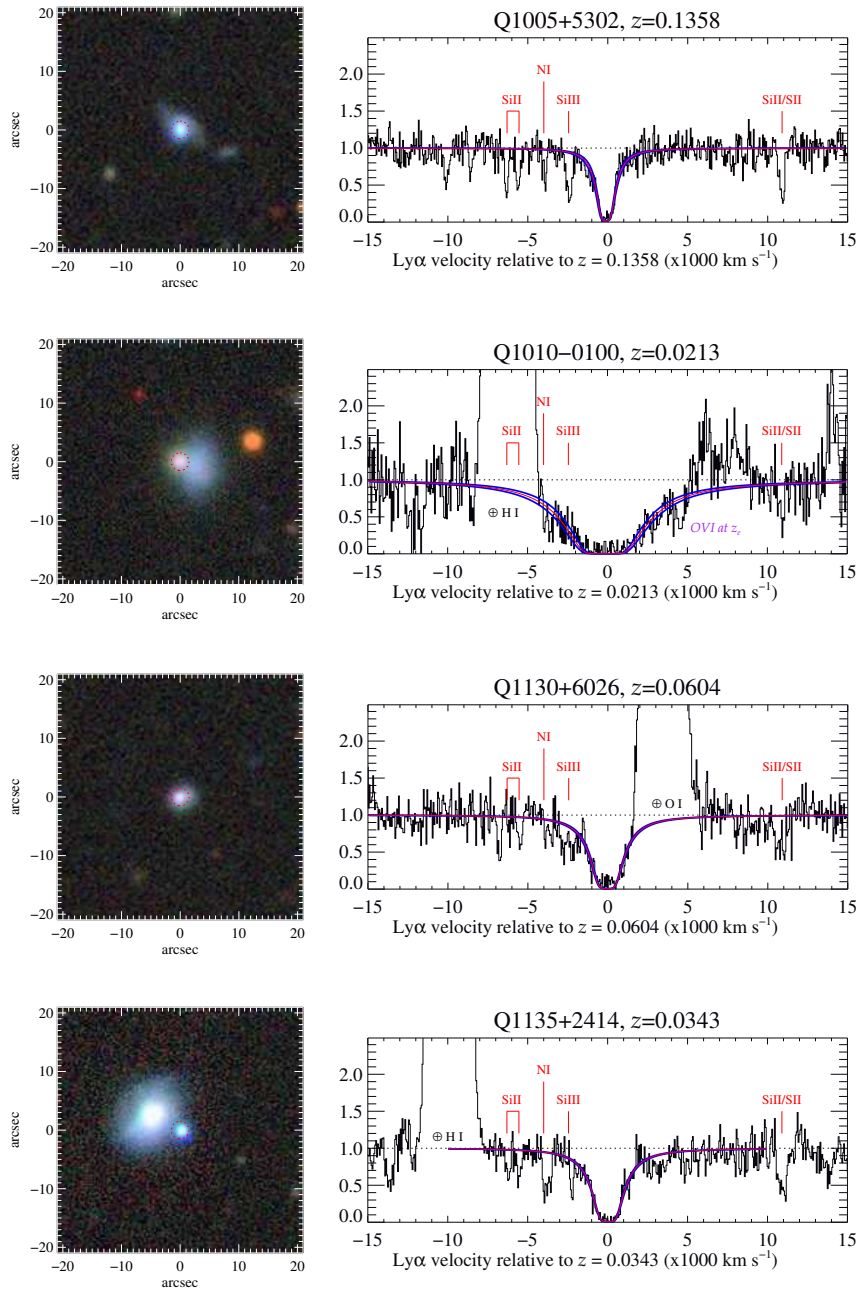

**Figure 1.** Four of the quasar-galaxy pairs in our sample. The images show $grz$-band cutouts of each of the quasar-galaxy pairs, taken from DR9 of the Dark Energy Camera Legacy Survey (DECaLS) (Dey et al 2019). Each image is centered on the co-ordinates of the quasar, and a red dotted circle indicates the area of the 3 arcsec diameter fibers used by SDSS to observe the quasar and galaxy. Also shown are portions of the G140L spectra of the quasars showing the strong Ly$\alpha$ absorption detected in the rest frame of the foreground galaxy. Voigt profile fits to the Ly$\alpha$ lines are shown as red lines, and the associated errors in the fits as purple lines. Geocoronal emission features and broad emission lines from the quasars are indicated. Metal absorption lines of Si II, Si III, N I and S II from the foreground galaxy which are close to the wavelength of Ly$\alpha$ are also indicated, but full details of the metal line absorption are given in Table 2.

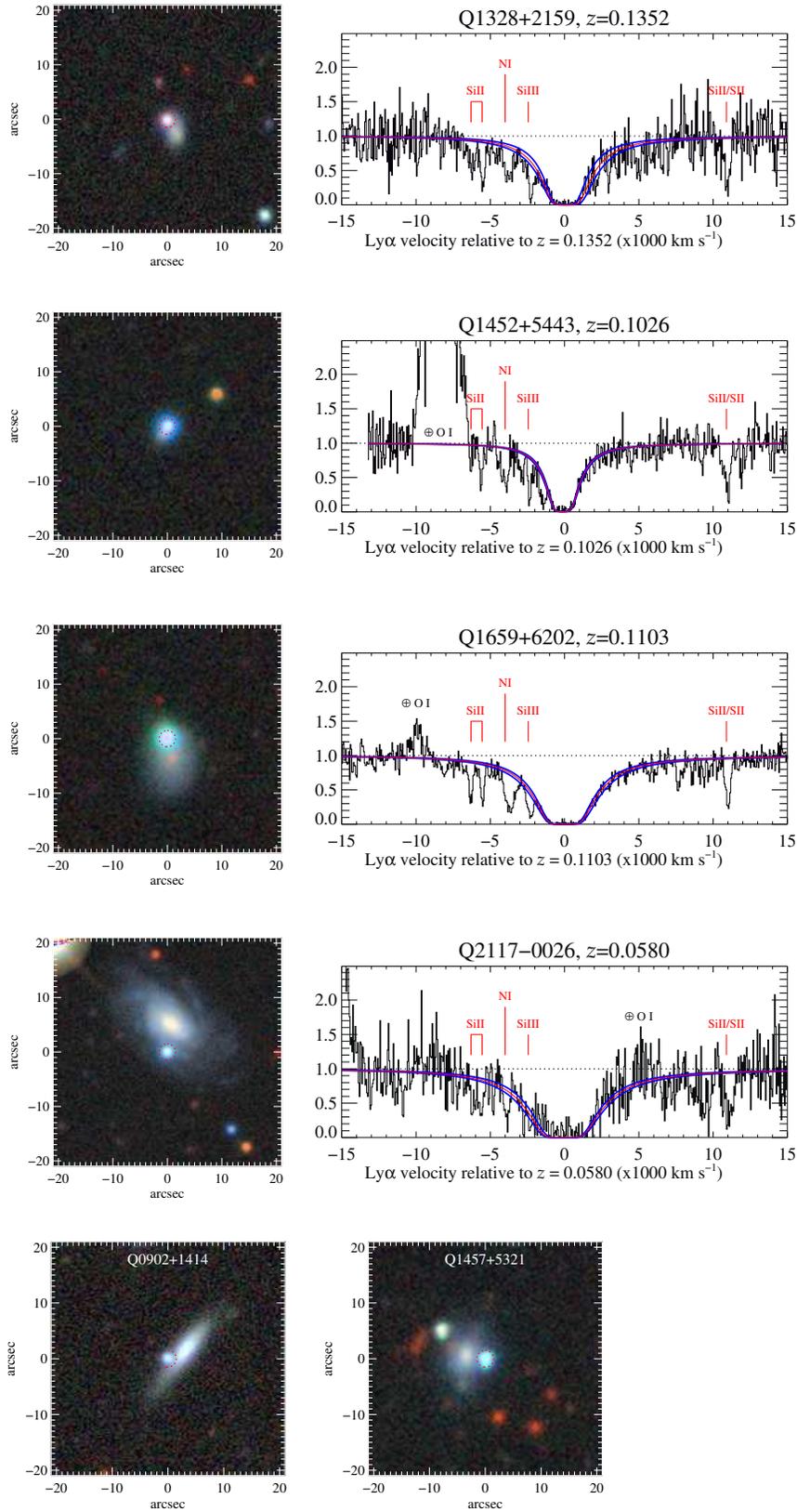

**Figure 2.** Same as Fig. 1, for our remaining quasar-galaxy pairs. No G140L spectra are shown for the two targets for which no detectable flux exists near the Lyα absorption owing to a higher redshift Lyman limit system.

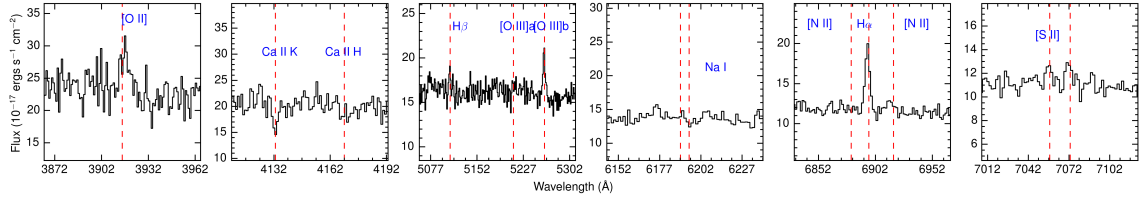
(a) SDSS spectra of Q0902+1414.

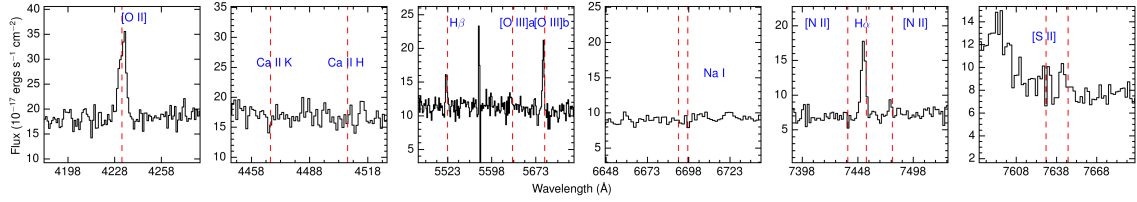
(b) SDSS spectra of Q1005+5302.

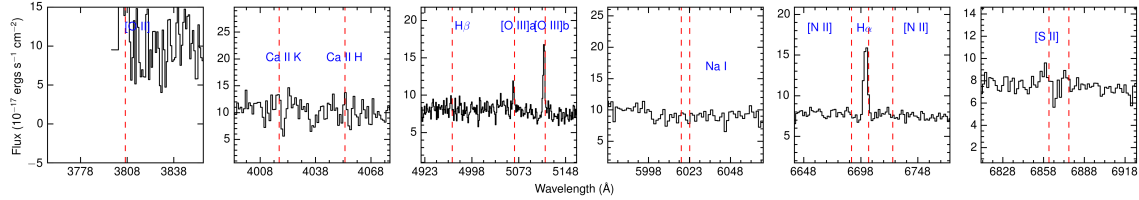
(c) SDSS spectra of Q1010-0100.

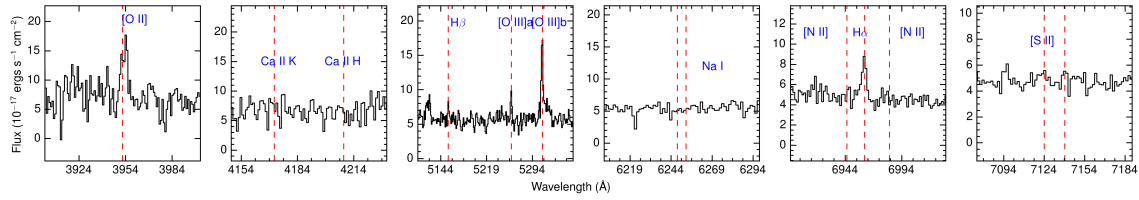
(d) SDSS spectra of Q1130+6026.

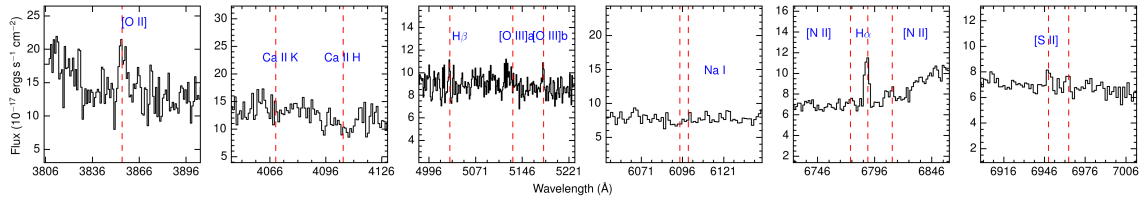
(e) SDSS spectra of Q1135+2414.

**Figure 3.** SDSS spectra of five of our target quasars. Dashed red lines mark the expected wavelengths of the nebular emission lines and the Ca II, Na I absorption lines from the foreground galaxies.

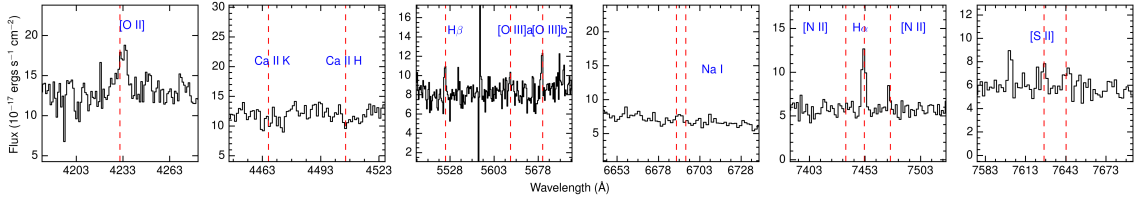

(a) SDSS spectra of Q1328+2159.

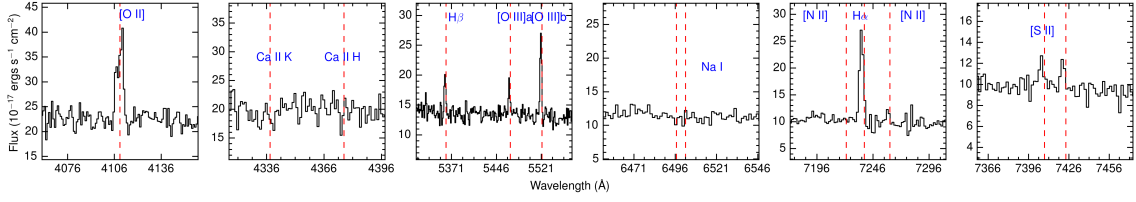

(b) SDSS spectra of Q1452+5443.

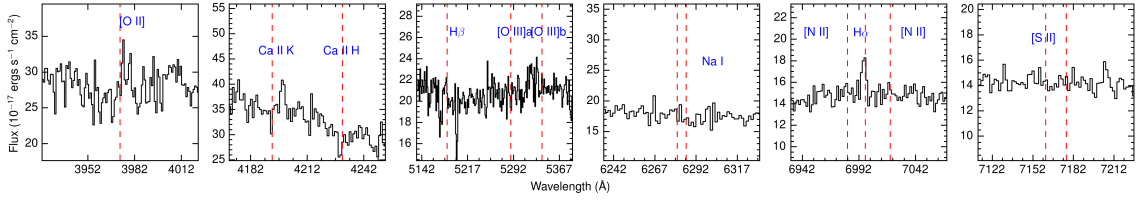

(c) SDSS spectra of Q1457+5321.

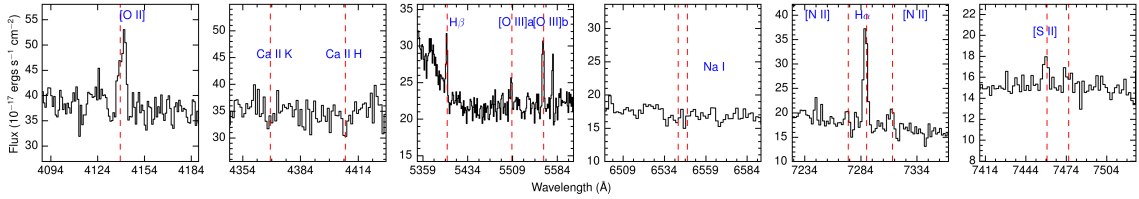

(d) SDSS spectra of Q1659+6202.

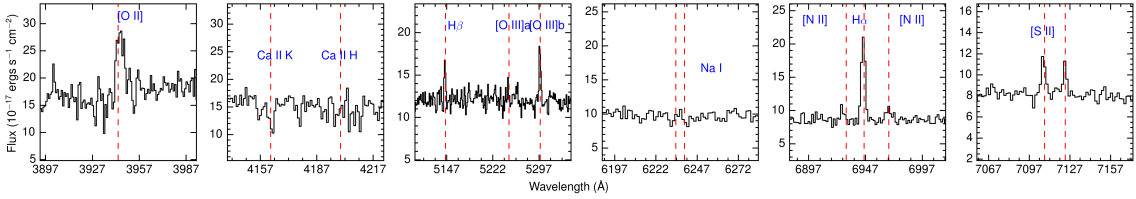

(e) SDSS spectra of Q2117-0026.

**Figure 4.** Same as Fig. 3, for our remaining target quasars.

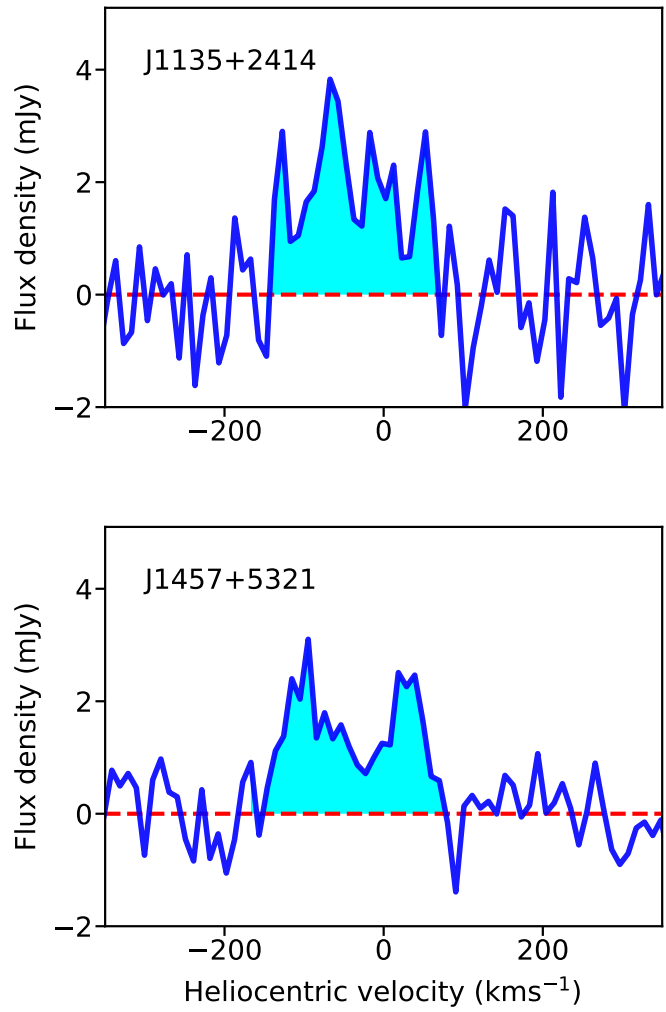

**Figure 5.** GBT spectra for two of our GOTOQs showing detections of H I 21-cm emission (highlighted in cyan). The velocities are with respect to $z=0.0343$ and $z=0.0660$, respectively, for Q11135+2414 and Q1457+5321.

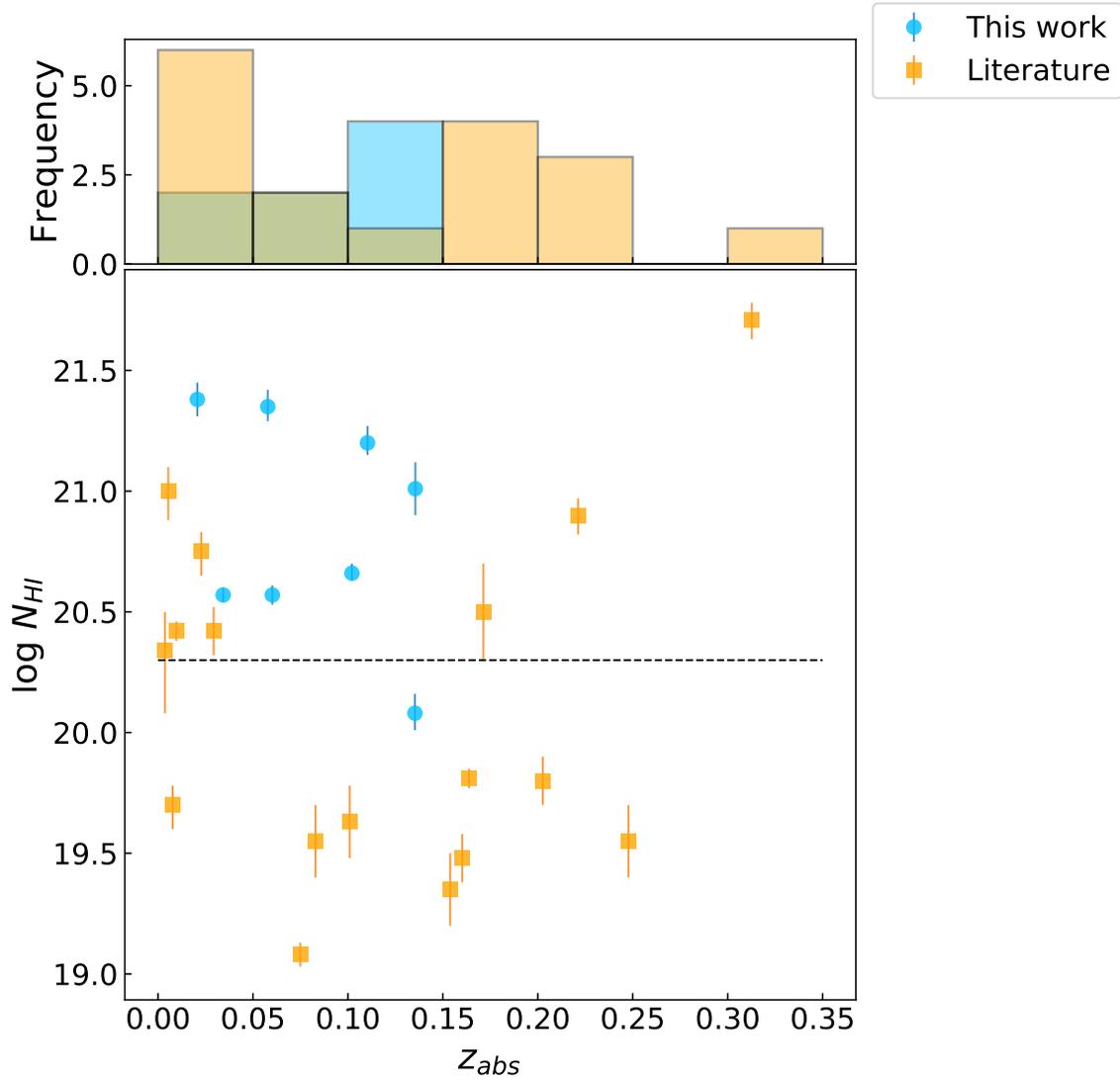

**Figure 6.** Top: Redshift distribution of absorbers from this study and those with redshifts $z<0.35$ from the literature. Bottom: H I column density vs. absorption redshift for absorbers from our study and those with redshifts $z<0.35$ from the literature. For reference, the horizontal dashed line shows the boundary between DLAs and sub-DLAs.

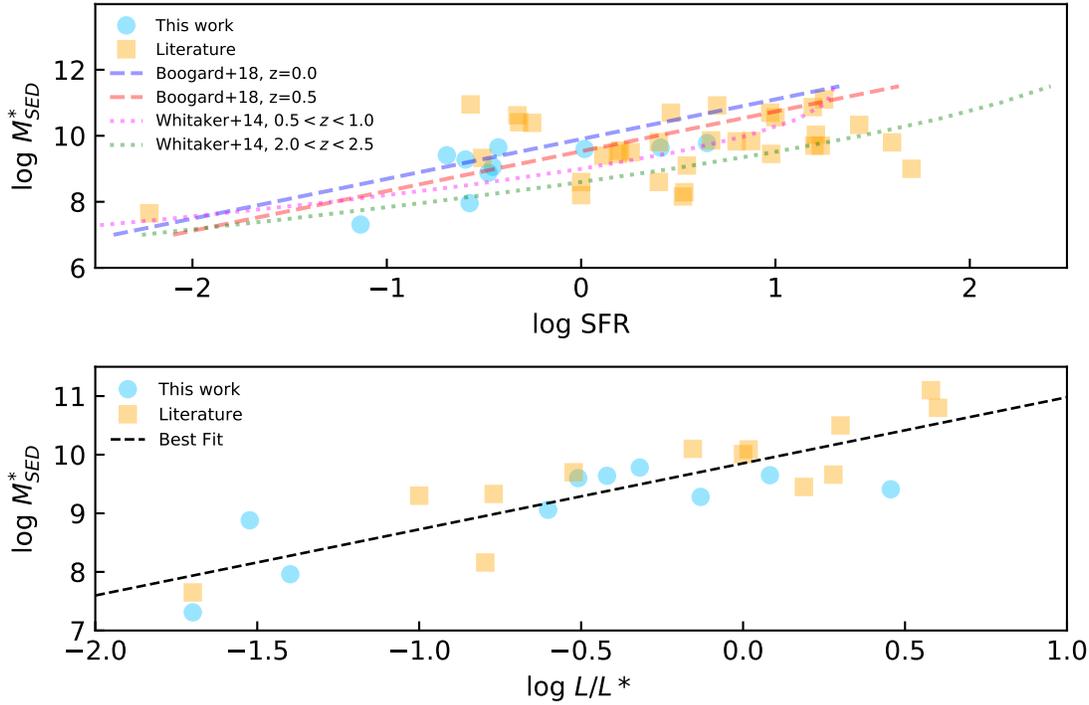

**Figure 7.** Top: Stellar mass based on SED fitting vs. SFR measurements for galaxies associated with DLA/sub-DLAs. Cyan circles show GOTOQs from our sample, while orange squares show absorber-galaxy pairs from the literature. Dashed and dotted curves show, for comparison, the SFMS from Boogard et al. (2018) and Whitaker et al. (2014) at a variety of redshifts. Bottom: Stellar mass based on SED fitting vs. luminosity for galaxies associated with DLA/sub-DLAs. Cyan circles show GOTOQs from our sample, while orange squares show absorber-galaxy pairs from the literature. In both panels, some orange squares appear darker than others because of the overlap of multiple data points with similar values. Dashed black line shows our best-fit based on linear regression. See text for more details.

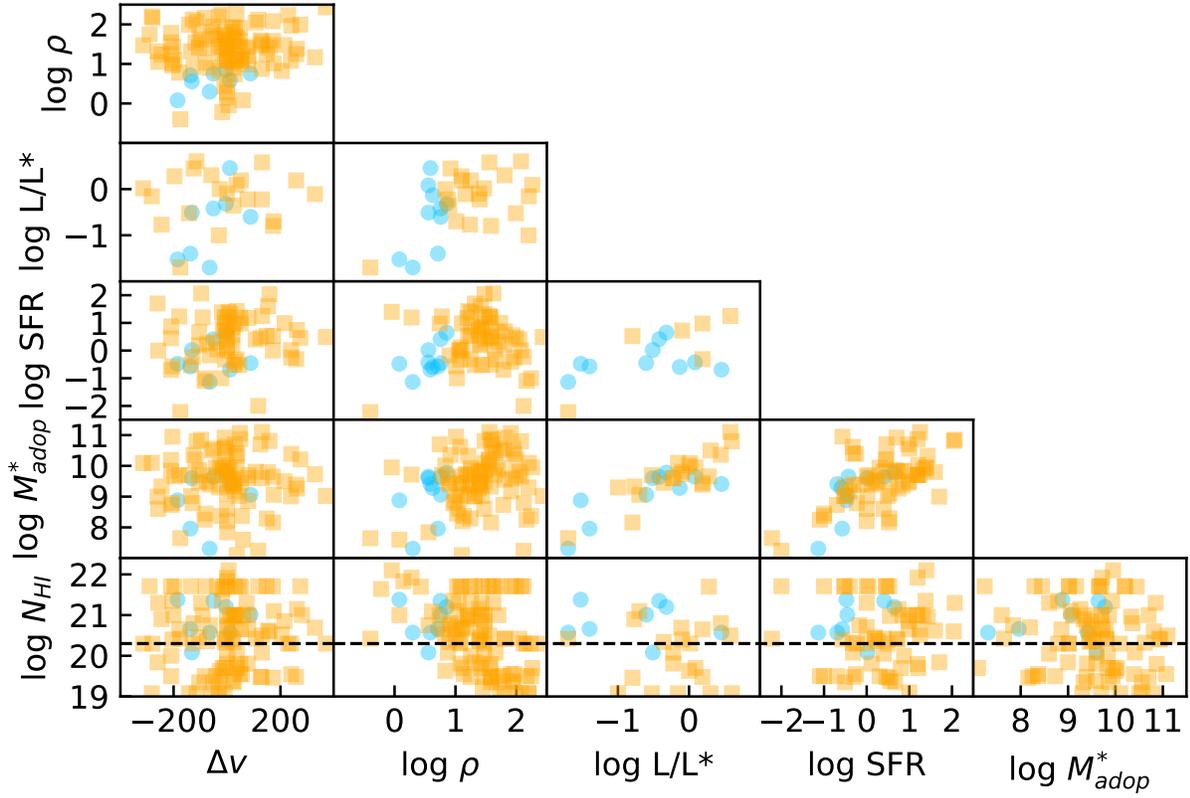

**Figure 8.** Various properties of galaxies associated with DLA/sub-DLA absorbers plotted against each other. Cyan circles show GOTOQs from our sample, while orange squares show absorber-galaxy pairs from the literature. Some orange squares appear darker than others because of the overlap of multiple data points with similar values. Dashed black lines in the lowest panels denote the separation in H I column density between DLAs and sub-DLAs. See text for more details.

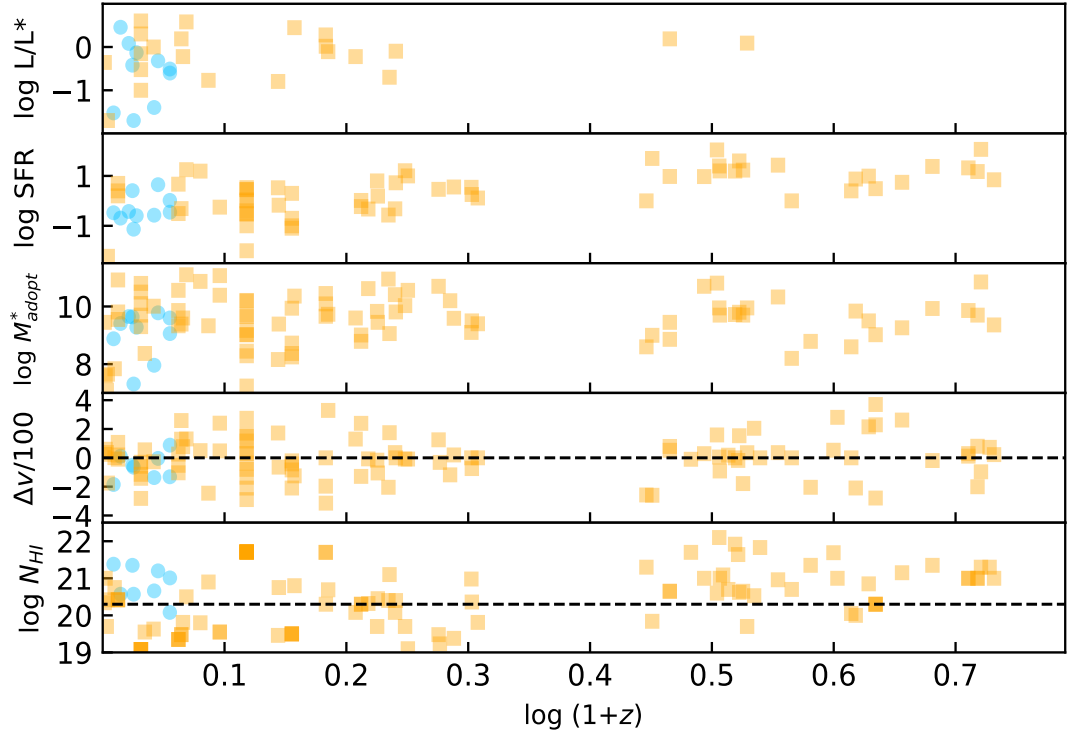

**Figure 9.** Trends in redshift of various galaxy and absorber properties. The panels show the luminosity, star formation rate, stellar mass of the galaxy, velocity offset of the absorber with respect to the galaxy, and the H I column density of the absorber plotted vs. redshift. Cyan circles show GOTOQs from our sample, while orange squares show galaxies from the literature. Dashed black line in the fourth panel from the top separates the cases of absorbers at higher redshifts with respect to the galaxies from those at lower redshifts. Dashed black line in the bottom panel denotes the separation in H I column density between DLAs and sub-DLAs. See text for more details.

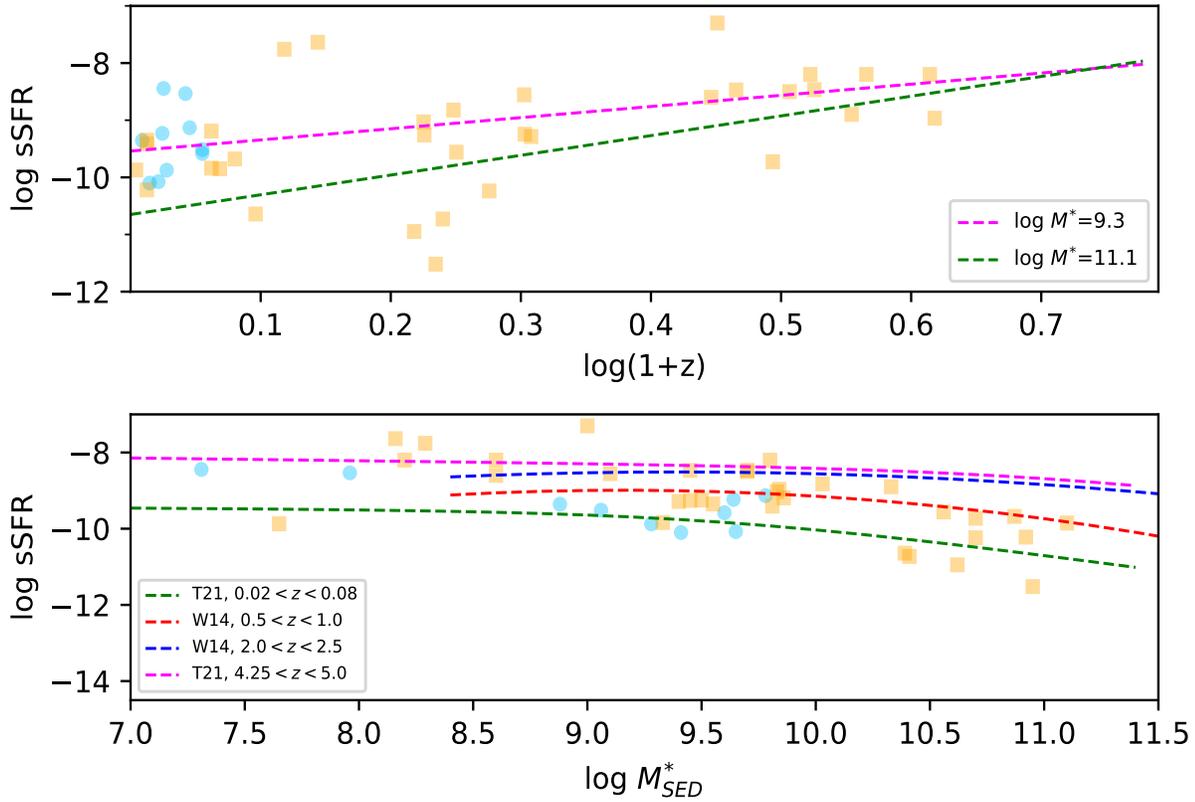

**Figure 10.** Specific star formation rate plotted vs. redshift (top panel), and vs. stellar mass (bottom panel) for DLA/sub-DLA galaxies that have stellar mass estimates from SED fitting. Cyan circles show GOTOQs from our sample, while orange squares show galaxies from the literature. Dashed curves in the upper panel show relations for star-forming galaxies for different stellar masses from Whitaker et al. (2014). Dashed curves in the lower panel show relations for star-forming galaxies for different redshift ranges from Whitaker et al. (2014) and Thorne et al. (2021). See text for more details.

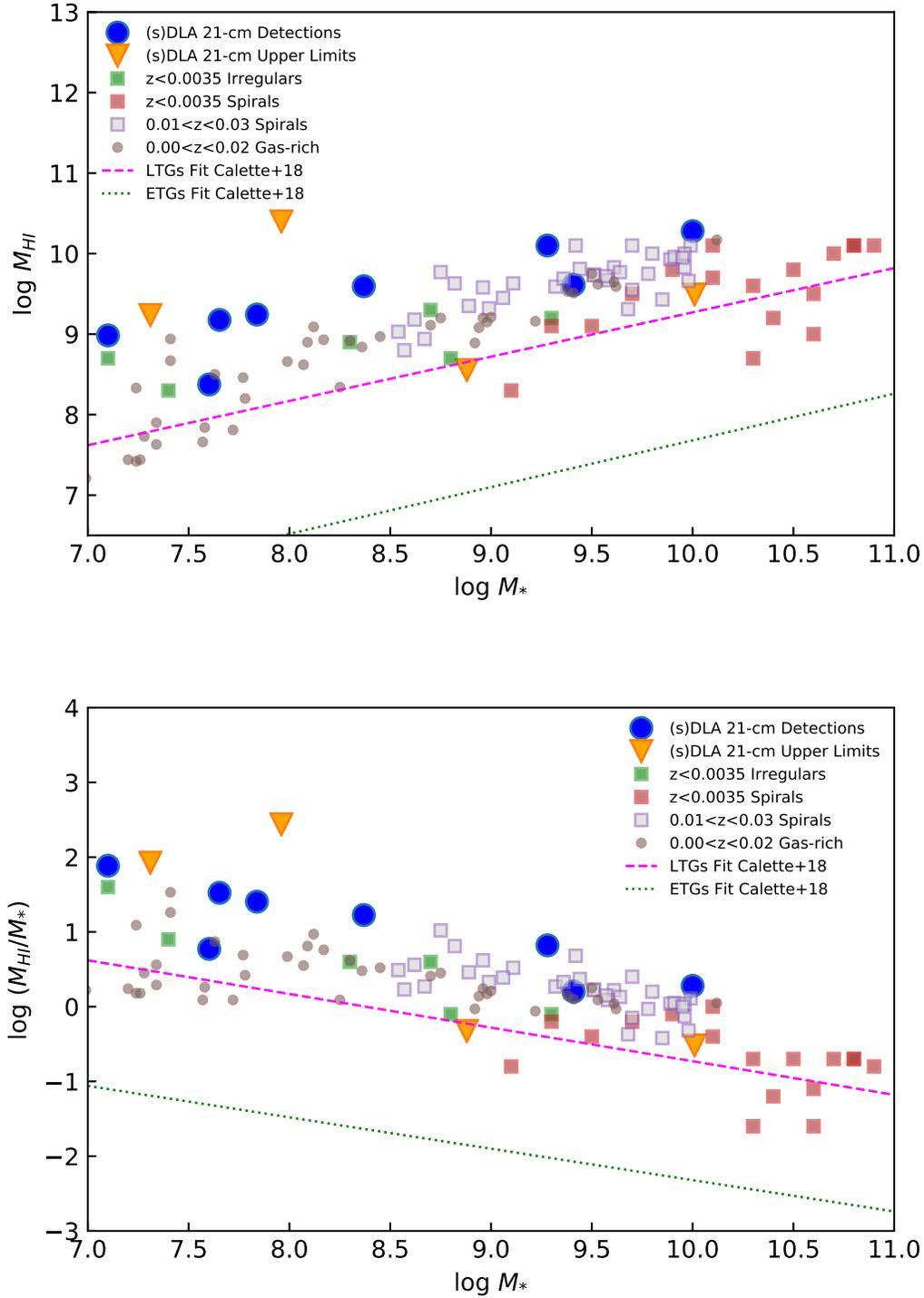

**Figure 11.** Top (a): H I gas mass vs. stellar mass for galaxies associated with DLA/sub-DLAs searched for in 21-cm emission from our sample and the literature. Other data points show H I and stellar masses of local spirals and irregulars, including gas-rich galaxies. Dashed and dotted lines show, respectively, the best-fitting trends found by (Calette et al. 2018) for local late-type galaxies and early-type galaxies. Bottom (b): Ratio of H I gas mass to stellar mass, plotted vs. stellar mass for galaxies associated with DLAs searched for in 21-cm emission from our sample and the literature. Symbols and lines have the same meaning as in the top panel. See text for more details.

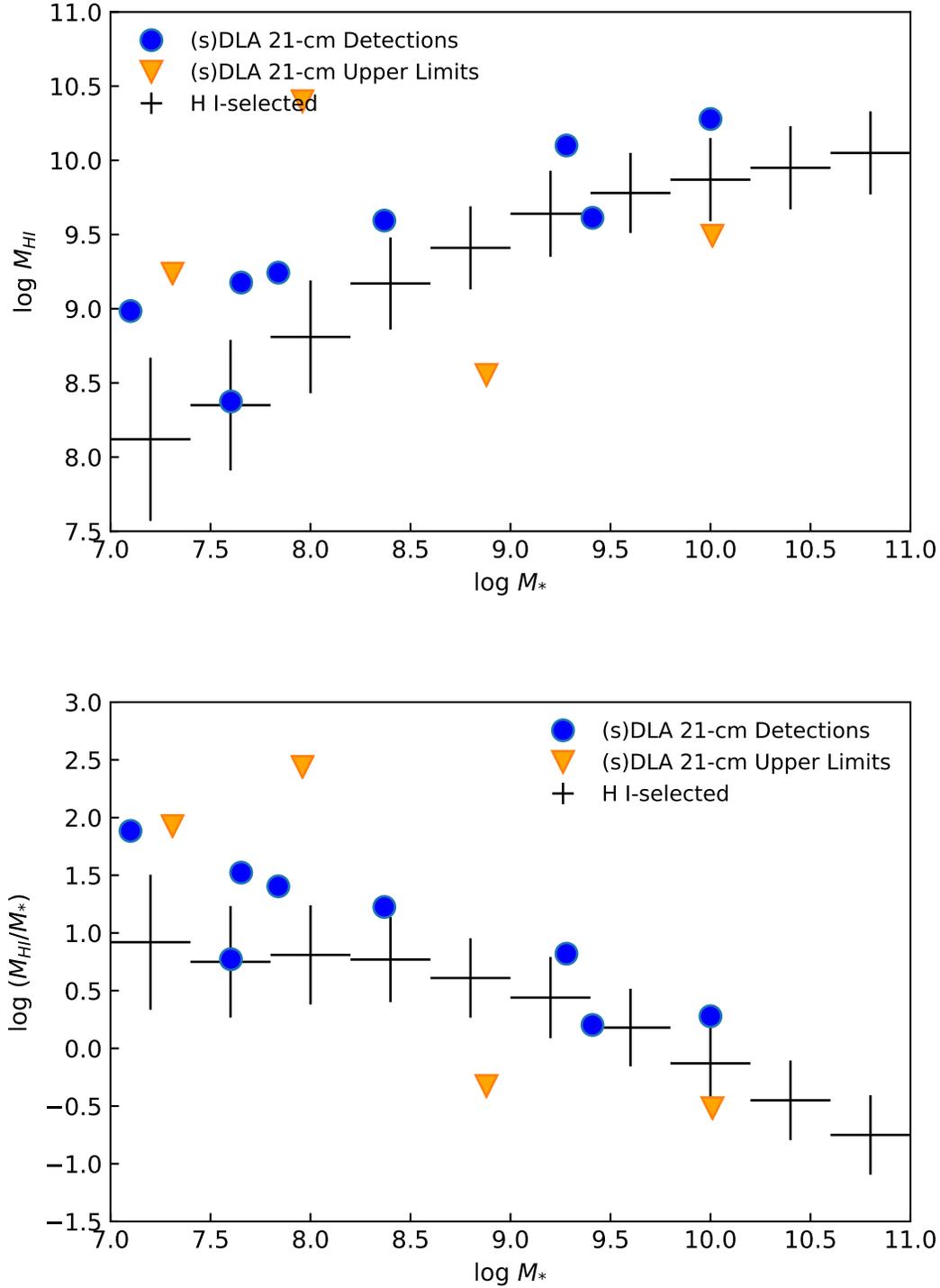

**Figure 12.** Top (a): H I gas mass vs. stellar mass for galaxies associated with DLAs searched for in 21-cm emission from our sample and the literature. Bottom (b): Ratio of H I gas mass to stellar mass, plotted vs. stellar mass for galaxies associated with DLAs searched for in 21-cm emission from our sample and the literature. Also shown in both panels are the corresponding relations (median values and $\pm 1\sigma$ uncertainties) deduced for galaxies selected by H I emission, based on 9,153 ALFALFA galaxies with SDSS spectra and stellar masses (Maddox et al. 2015). See text for more details.

Table 3. COS absorption characteristics

| Quasar | log $N_{HI}$ (cm$^{-2}$) | $\Delta v^1$ (km s$^{-1}$) | $A^3_{COS}$ (pkpc$^2$) | $A^4_{SDSS}$ (pkpc$^2$) |
|---|---|---|---|---|
| Q0902+1414[2] | – | – | 5 | 7 |
| Q1005+5302 | $20.08^{+0.08}_{-0.07}$ | $-131.5^{+9.8}_{-7.9}$ | 28 | 41 |
| Q1010-0100 | $21.38^{+0.07}_{-0.07}$ | $-185.0^{+29.0}_{-29.0}$ | 1 | 2 |
| Q1130+6026 | $20.57^{+0.04}_{-0.04}$ | $-65.0^{+8.0}_{-8.0}$ | 7 | 10 |
| Q1135+2414 | $20.57^{+0.03}_{-0.03}$ | $11.0^{+8.0}_{-8.0}$ | 3 | 3 |
| Q1328+2159 | $21.01^{+0.11}_{-0.11}$ | $88.0^{+61.0}_{-46.0}$ | 28 | 41 |
| Q1452+5443 | $20.66^{+0.04}_{-0.03}$ | $-137.0^{+6.0}_{-6.0}$ | 18 | 25 |
| Q1457+5321[2] | – | – | 8 | 12 |
| Q1659+6202 | $21.20^{+0.07}_{-0.05}$ | $-4.0^{+10.0}_{-10.0}$ | 20 | 30 |
| Q2117-0026 | $21.35^{+0.07}_{-0.06}$ | $-51.0^{+48.0}_{-37.0}$ | 6 | 9 |

1. Velocity offset of the absorption system relative to the galaxy emission redshift (determined from nebular emission lines in SDSS spectra; see Figures 3 and 4). 2. Two sight lines from our sample had no detectable flux from the quasar, indicating the presence of higher-redshift LLSs. See text for more details. 3. The physical area probed by the COS aperture at the galaxy redshift. 4. The physical area probed by the SDSS aperture at the galaxy redshift.

Table 4. H I 21-cm Emission Measurements

| quasar | $z_{gal}$ | $z_{abs}$ | Run ID | RMS (mJy) | $M_{\rm HI}$ ($10^9 M_\odot$) |
|---|---|---|---|---|---|
| J1010-0100 | 0.0213 | 0.02067 | 3 | 1.28 | <0.36 |
| J1130+6026 | 0.0604 | 0.06017 | 1, 2 | 0.73 | <1.72 |
| J1135+2414 | 0.0343 | 0.03434 | 4 | 0.83 | 4.1 |
| J1452+5443 | 0.1026 | 0.10210 | 4, 5, 7 | 3.64 | <25.0 |
| J1457+5321 | 0.0660 | ...[1] | 6 | 0.56 | 12.6 |

[1] DLA could not be covered due to the presence of a higher-redshift Lyman-limit system.

**Table 5.** Results of Correlation Tests between galaxy and absorber parameters

| Parameters | No. of objects[1] | $r_S^2$ | $P^3$ | Fig., panel[4] |
|---|---|---|---|---|
| Significant Correlations: | | | | |
| $z$, $\log N_{\rm HI}$ | 83 | 0.324 | $2.83 \times 10^{-3}$ | 9, 1, 5 |
| $z$, $\log$ SFR | 73 | 0.607 | $1.25 \times 10^{-8}$ | 9, 1, 2 |
| | | | | |
| $\log N_{\rm HI}$, $\log \rho$ | 111 | -0.413 | $6.79 \times 10^{-6}$ | 8, 2, 5 |
| $\log N_{\rm HI}$, $\log M^*_{\rm SED}$ | 60 | -0.342 | $7.52 \times 10^{-3}$ | ... |
| | | | | |
| $\log L/L_*$, $\log M^*_{\rm adopt}$ | 32 | 0.752 | $7.04 \times 10^{-7}$ | 8, 3, 4 |
| $\log L/L_*$, $\log M^*_{\rm SED}$ | 23 | 0.791 | $6.95 \times 10^{-6}$ | 7, 1, 2 |
| $\log L/L_*$, $\log$ SFR | 16 | 0.526 | $3.65 \times 10^{-2}$ | 8, 3, 3 |
| | | | | |
| $\log \rho$, $\log M^*_{\rm adopt}$ | 98 | 0.276 | $5.99 \times 10^{-3}$ | 8, 2, 4 |
| $\log \rho$, $\log M^*_{\rm SED}$ | 61 | 0.528 | $1.23 \times 10^{-5}$ | ... |
| | | | | |
| $\log$ SFR, $\log M^*_{\rm adopt}$ | 73 | 0.575 | $1.05 \times 10^{-7}$ | 8, 4, 4 |
| $\log$ SFR, $\log M^*_{\rm SED}$ | 43 | 0.445 | $2.81 \times 10^{-3}$ | 7, 1, 1 |
| | | | | |
| $\log$ sSFR, $\log (1+z)$ | 43 | 0.452 | $2.33 \times 10^{-3}$ | 10, 1, 1 |
| $\log$ sSFR, $\log M^*_{\rm SED}$ | 43 | -0.548 | $1.42 \times 10^{-4}$ | 10, 1, 2 |
| | | | | |
| Insignificant Correlations: | | | | |
| $z$, $\log L/L_*$ | 32 | 0.269 | 0.137 | 9, 1, 1 |
| $z$, $\log \rho$ | 113 | 0.0302 | 0.751 | ... |
| $z$, $\log M^*_{\rm adopt}$ | 99 | 0.147 | 0.147 | 9, 1, 3 |
| $z$, $\log M^*_{\rm SED}$ | 62 | 0.158 | 0.220 | ... |
| $z$, $\Delta v$ | 113 | 0.0991 | 0.296 | 9, 1, 4 |
| $\log N_{\rm HI}$, $\log M^*_{\rm adopt}$ | 97 | -0.173 | 0.0901 | 8, 5, 5 |
| $\log N_{\rm HI}$, $\log L/L_*$ | 30 | -0.178 | 0.346 | 8, 3, 5 |
| $\log N_{\rm HI}$, $\log$ SFR | 71 | 0.0648 | 0.592 | 8, 4, 5 |
| $\log N_{\rm HI}$, $\Delta v$ | 113 | 0.0413 | 0.664 | 8, 1, 5 |
| $\log L/L_*$, $\log \rho$ | 32 | 0.347 | 0.0514 | 8, 2, 2 |
| $\log L/L_*$, $\Delta v$ | 30 | 0.171 | 0.365 | 8, 1, 2 |
| $\log M^*_{\rm adopt}$, $\Delta v$ | 97 | -0.0125 | 0.903 | 8, 1, 4 |
| $\log \rho$, $\log$ SFR | 72 | 0.0592 | 0.622 | 8, 2, 3 |
| $\log \rho$, $\Delta v$ | 111 | 0.0862 | 0.368 | 8, 1, 1 |
| $\log$ SFR, $\Delta v$ | 71 | 0.183 | 0.126 | 8, 1, 3 |
| $\log M^*_{\rm SED}$, $\Delta v$ | 60 | 0.0163 | 0.902 | ... |

1. No. of galaxy/absorber pairs for which both parameters are available. 2. Spearman rank order correlation coefficient. 3. Probability that the obsered value of $r_S$ could arise purely by chance. 4. Figure and panel number showing the trend between the particular parameters. First index denotes the figure number, the second index denotes the column number of the panel in that figure from the left, and the third index denotes the row number of the panel from the top of the entire figure. For Figs. 7, 9, and 10, which contain only one column, we denote the column number as 1.